\documentclass[
 reprint,
 citeautoscript,
 amsmath,amssymb,
 aps,
 prb,
]{revtex4-2}

\usepackage{graphicx}
\usepackage{dcolumn}
\usepackage{bm}
\usepackage{xcolor}
\usepackage[english]{babel}

\makeatletter
\let\ORIbbl@fixname\bbl@fixname
\def\bbl@fixname#1{%
  \@ifundefined{languagealias@\expandafter\string#1}
    {\ORIbbl@fixname#1}
    {\edef\languagename{\@nameuse{languagealias@#1}}}%
}
\newcommand{\definelanguagealias}[2]{%
  \@namedef{languagealias@#1}{#2}%
}
\makeatother

\definelanguagealias{en}{english}
\definelanguagealias{EN}{english}

\newcommand{\FS}[1]{\left\langle #1 \right\rangle_{\scriptscriptstyle\mathrm{FS}}}
\newcommand{\Eq}[1]{Eq.~(\ref{#1})} 
\newcommand{\kb}{k_\mathrm{B}}
\newcommand{\tc}{T_\mathrm{c}}

\newcommand{\Ib}{I_\mathrm{b}}
\newcommand{\Ia}{I^\mathrm{a}}
\newcommand{\Ile}{I^\mathrm{le}}
\newcommand{\RR}{\mathbf{R}}
\newcommand{\vF}{\mathbf{v}_\mathrm{F}}
\newcommand{\pF}{\mathbf{p}_\mathrm{F}}
\newcommand{\ps}{\mathbf{p}_\mathrm{s}}
\newcommand{\mfp}{\ell}
\newcommand{\NF}{\mathcal{N}_\mathrm{F}}
\newcommand{\Ac}{\mathcal{A}}
\newcommand{\Vc}{V_\mathrm{c}}
\newcommand{\fL}{f_\mathrm{L}}
\newcommand{\fT}{f_\mathrm{T}}
\newcommand{\fe}{f_\mathrm{e}}

\begin{document}

\title{Quasiclassical theory of charge transport across mesoscopic normal metal-superconducting heterostructures with current conservation}

\author{Kevin Marc Seja}
\author{Tomas L\"ofwander}
\affiliation{Department of Microtechnology and Nanoscience - MC2,
Chalmers University of Technology,
SE-41296 G\"oteborg, Sweden}

\date{\today}

\begin{abstract}
We consider the steady-state nonequilibrium behavior of mesoscopic superconducting wires connected to normal-metal reservoirs. Going beyond the diffusive limit, we utilize the quasiclassical theory and perform a self-consistent calculation that guarantees current conservation through the entire system. Going from the ballistic to the diffusive limit, we investigate several crucial phenomena such as charge imbalance, momentum-resolved nonequilbrium distributions, and the current-to-superflow conversion. Connecting to earlier results for the diffusive case, we find that superconductivity can break down at a critical bias voltage $V_\mathrm{c}$. We find that $V_\mathrm{c}$  generally increases as the interface transparency is reduced, while the dependence on the mean-free path is non-monotonous. We discuss the key differences of the ballistic and semi-ballistic regimes to the fully diffusive case.

\end{abstract}

\maketitle

\section{Introduction}

The non-equilibrium distribution of quasiparticles in hybrid nanostructures has for a long time been an important topic within the field of superconductivity \cite{clarke_experimental_1972,tinkham_theory_1972,Schmid1975Jul,artemenko_electric_1979,yagi2006,hubler_charge_2010,arutyunov_relaxation_2018}. The interest arises naturally due to the wide range of applications of superconducting devices in sensing, metrology, thermometry and refridgeration at the nanoscale, as well as high-speed and quantum computing \cite{giazotto_opportunities_2006,pekola_single-electron_2013,anders_european_2010,fornieri_towards_2017,wendin_quantum_2017}. The non-equilibrium distribution may be a consequence of device operation, but in many cases unwanted, or poisonous, quasiparticles may limit device performance \cite{de_visser_number_2011,Catelani2021}. Also within the field of superconducting spintronics in ferromagnetic-superconducting hybrid structures, the spin-dependent distributions have been investigated \cite{hubler_long-range_2012,quay_spin_2013,beckmann_spin_2016,eschrig_spin-polarized_2015,Bergeret2018}. Recent advances in nano-device fabrication involving unconventional superconductors \cite{nawaz_microwave_2013,andersson_fabrication_2020,trabaldo_properties_2020} also raise questions about non-equilibrium effects in such systems. One fundamental problem, from the perspective of theory, is that current conservation in hybrid structures is only guaranteed if self-consistency of the relevant self-energies is taken into account.

In many cases the neglect of current conservation is a good approximation, for instance due to a very small perturbation that enables a linear response approach. Although also in this case vertex corrections can be important to include. For the pinhole geometry between a normal metal and a superconductor, the dilution of the current flowing through the pinhole into the reservoirs leads to only small corrections if self-consistency is taken fully into account \cite{Blonder1982Apr}. 

In other cases current conservation is more important. For instance in nanoscale to mesoscale devices, where the non-equilibrium distribution function and the presence of current severely affect the superconducting order parameter and other self-energies \cite{martin_self-consistent_1995,canizares_self-consistent_1995,canizares_absence_1996,sanchez-canizares_current-conserving_1996,sanchez-canizares_self-consistent_1997,sols_conductances_1999,sanchez-canizares_self-consistent_2001,boogaard_resistance_2004,Keizer2006Apr,vercruyssen_evanescent_2012}. Then it becomes crucial to take into account the conversion between quasiparticle current and superflow through Andreev processes and the spatial variation of the order parameter on the length-scale of the superconducting coherence length.

In this paper, our aim is to establish an efficient computational strategy for calculating conserved current flow in mesoscopic hybrid structures including superconducting and normal-metal elements with mean free paths $\mfp$ ranging from the clean limit $\mfp\gg\xi_0$ to the dirty limit $\mfp\ll\xi_0$, where $\xi_0=\hbar v_\mathrm{F}/2\pi\tc$ is the clean-limit superconducting coherence length defined in terms of Planck's constant $\hbar$, the Fermi velocity in the normal state $v_\mathrm{F}$, and the superconducting transition temperature $\tc$. We utilize the quasiclassical theory of superconductivity pioneered by Eilenberger, Larkin, and Ovchinnikov \cite{Eilenberger1968Apr,Larkin1969}. Within this Green's function method we take self-consistency of the superconducting order parameter as well as impurity self-energies into account. We limit ourselves to nanoscale devices where the inelastic mean free path $\mfp_\mathrm{in}$ is larger than the device size, although this is not a limitation of the theory \cite{rainer_strong-coupling_1995}. In addition, we focus on heterostructures with only one superconducting element. In this case the non-equilibrium state is stationary \cite{snyman_bistability_2009}.

In the diffusive limit, when the mean free path $\mfp$ is much smaller than the superconducting coherence length, $\ell\ll\xi_0$, the Usadel approximation can be made \cite{usadel_generalized_1970,belzig_quasiclassical_1999}. The Usadel equations, being diffusion-type equations, may be solved numerically for instance with a finite element method \cite{amundsen_general_2016}, or by implementing the equations of Nazarov's circuit theory \cite{nazarov_novel_1999}. Going beyond the diffusive limit, it is necessary to go back to the more general Eilenberger-Larkin-Ovchinnikov formulation.

First non-equilibrium calculations including current conservation were made in the 1990s \cite{martin_self-consistent_1995,canizares_self-consistent_1995,canizares_absence_1996,sanchez-canizares_current-conserving_1996,sanchez-canizares_self-consistent_1997,sols_conductances_1999,sanchez-canizares_self-consistent_2001}. In the work by Sols and Sánchez-Cañizares \cite{canizares_self-consistent_1995,canizares_absence_1996,sanchez-canizares_current-conserving_1996,sanchez-canizares_self-consistent_1997,sols_conductances_1999} a scattering approach was utilized with an approximate scheme of asymptotic self-consistency in the clean limit. The importance of spectral rearrangement in the form of Doppler shifts of quasiparticle states in the presence of superflow was pointed out. Later \cite{sanchez-canizares_self-consistent_2001}, impurity scattering was also included but only very small systems of the order of three coherence lengths were studied. 

This paper is organized as follows. In Section II we outline the model assumptions, give details of the quasiclassical theoretical framework, and outline our computational strategies. In section III we report results on a range of normal metal-superconducting heterostructures, sketched in Fig.~\ref{fig:sketch}. Section IV summarizes the paper.

\section{Model and Methods}

\begin{figure}
    \centering
    \includegraphics[width=\columnwidth]{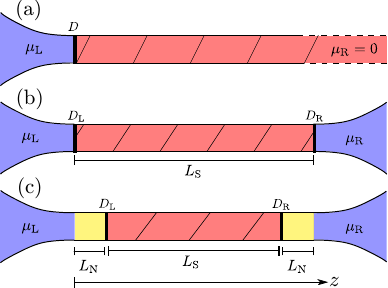}
    \caption{
    Sketch of the various device set-ups. (a) IS system. A superconductor (red, grated) is connected on one end to a normal-metal reservoir (blue) via an insulating barrier (black) of transparency $D$. The reservoir is at a chemical potential $\mu_\mathrm{L}$, while the superconductor is assumed to be grounded on the right-hand side.
    (b) ISI system. The central superconducting region (red, grated) of length $L_\mathrm{S}$ is connected to normal-metal reservoirs (blue) on both ends via insulating barriers (black). The transparencies $D_\mathrm{L,R}$ do not have to be symmetric, in this case $|\mu_\mathrm{L}| \neq |\mu_\mathrm{R}|$.
    (c) NISIN system. The central superconducting region (red, grated) of length $L_\mathrm{S}$ is connected via insulating barriers (black) to a left and right reservoir (blue) via a finite piece of normal wire (yellow) of length $L_\mathrm{N}$ rather then directly as in case (b). In these normal-metal wires the proximity effect is taken into account. The spatial coordinate along the wire main axis is denoted $z$.}
    \label{fig:sketch}
\end{figure}

In this paper we shall study a number of normal metal-superconducting devices with steady-state current flow between source and drain contacts, see Fig.~\ref{fig:sketch}. The superconductor is considered to be at ground potential in equilibrium, while in non-equilibrium we also compute the local electrochemical potential $\phi(z)$ self-consistently. This potential describes the local difference in potential between quasiparticles and the condensate. Our study is thus directly related to the charge imbalance problem studied for a long time and recently reviewed in Ref.~\cite{arutyunov_relaxation_2018}. Here we revisit this problem and compute all quantities self-consistently to guarantee current conservation. Moreover, we go beyond the diffusive approximation and cover the whole range of mean free paths from the clean to the dirty limits.

We shall consider a number of set-ups, starting in Section~\ref{subsec:IS} with a normal metal reservoir tunnel coupled to a superconductor which we assume continues to a grounded reservoir on the other side, see Fig.~\ref{fig:sketch}(a). This is in principle the usual set-up in the scattering approach to tunneling into a superconductor, pioneered for the ballistic case by Blonder-Tinkham-Klapwijk (BTK) in Ref.~\cite{Blonder1982Apr}. It has been extended and widely used also for unconventional superconductors, see, e.g., the reviews in Refs.~\cite{kashiwaya_tunnelling_2000,lofwander_andreev_2001}. The approach typically neglects self-consistency and current conservation, which is a good approximation when there is a Sharvin, or pinhole-like, contact to the superconductor that guarantees current dilution into the contacts. We are here interested in the corrections when this assumption is not strictly valid, similar to the situation in the experiments in Refs.~\cite{boogaard_resistance_2004,vercruyssen_evanescent_2012}.

We limit our study to small mesoscopic devices at a low temperature $T=0.01T_c$ with good contacts to the reservoirs. More specifically, the system size $L$ is assumed sufficiently small and the contacts to the reservoirs are assumed to be sufficiently good that the inelastic scattering time is the longest time-scale in the problem and for instance electron-phonon relaxation and recombination processes can be neglected \cite{Kaplan1976,Takane2006,Catelani2010}, see discussion below.
As the main consequence, we cannot properly describe charge imbalance relaxation when the applied voltage is sufficiently high that a considerable amount of quasiparticles are injected into continuum states in the superconductor in the setup shown in Fig.~\ref{fig:sketch}(a). The potential $\phi(z)$ becomes finite far into the bulk of the superconductor and charge imbalance does not decay. In this case, the setup in Fig.~\ref{fig:sketch}(a) becomes ill-defined in the sense that we can not calculate the distribution function far in the superconductor. We therefore consider the well-defined setup in Fig.~\ref{fig:sketch}(b), where there is a normal metal reservoir also on the right side. In this case we study charge imbalance self-consistently on distances smaller than the inelastic scattering length $L\ll\mfp_\mathrm{in}$.

Finally, we study the case when parts of the normal metal leads to the superconductor are influenced by the proximity effect, see Fig.~\ref{fig:sketch}(c). This leads to a series resistance that depends on both system size and elastic mean free path.

To justify that inelastic scattering can be neglected at low temperature and small device sizes, we should compare the electron-phonon scattering times, in particular the recombination time $\tau_r$, and the time for diffusion from one contact to the other, or dwell time in the device, $\tau_\mathrm{d}$.
Kaplan {\it et al.} \cite{Kaplan1976} showed that in a bulk superconductor with thermal distributions the electron-phonon recombination time is exponentially long at low temperature, $\tau_r\propto\tau_0 e^{\Delta/\kb T}$, where $\Delta$ is the superconducting gap, while the electron-phonon scattering time follows a power-law, $\tau_s\propto\tau_0(\Delta/T)^{7/2}$. The typical time-scale $\tau_0\sim 0.1-1$~ns for strong-coupling superconductors, but is as long as $\tau_0\sim 500$ ns in weak-coupling superconductors. When the quasiparticle distribution is not thermal, experiments show a wide range of low-temperature recombination time scales ranging form $10~\mu$s up to $\sim 1$~ms \cite{Barends2009,de_visser_number_2011,Catelani2021}.
Catelani {\it et al.} \cite{Catelani2010} showed that in the diffusive limit and at low voltage, the non-equilibrium population in a small superconducting device induced by quasiparticle injection from normal metal reservoirs decays even weaker through electron-phonon scattering processes if the quasiparticle life-time due to escape to reservoirs is small $\tau_d\ll \tau_0\sqrt{\Delta/T}e^{\Delta/T}$. In our studies below we shall see that the non-equilibrium occupation of continuum states is indeed very small, as in Ref.~\cite{Catelani2010}, and the moving condensate (finite superfluid momentum below) carries the current in the interior of the devices as a result of Andreev processes near the interfaces. We note in passing, that such processes do not lead to heat flowing into the superconductor \cite{Andreev1964}.
Let us compare the recombination time $\tau_r$ to the escape time to reservoirs $\tau_\mathrm{d}$.
In the normal state we use the Thouless energy $E_\mathrm{T}=\hbar D/L^2$ with the diffusion constant $D=\tfrac{1}{3}v_\mathrm{F}\ell$ to estimate $\tau_d=\hbar/E_\mathrm{T}\sim 20$ ns, assuming a typical Fermi velocity $v_\mathrm{F}\sim 10^6$ m/s, system size $L\sim 100\xi_0$, with the clean limit $\xi_0\sim 1~\mu$m, and mean free path $\ell\sim\xi_0$. Since we assume high transparencies of the barriers to the leads, the escape time is dominated by the diffusion time. Thus, in the normal state the escape time to reservoirs is much shorter than the electron-phonon scattering time, $\tau_d\ll\tau_0$. In the superconducting state the low group velocity of quasiparticle states leads to a reduction of the diffusion constant as $D\rightarrow D\sqrt{\varepsilon^2-\Delta^2}/|\varepsilon|$ and the escape time $\tau_d$ is enhanced at the gap edge. Experimentally it was measured to give approximately a factor of 10 reduction of the effective diffusion constant at low temperatures \cite{Ullom1998}. At the same time it should be compared with $\tau_r$ typically enhanced at low temperatures in the superconducting state by several orders of magnitude. In conclusion, we assume good contacts to reservoirs allowing for efficient Andreev processes, a weak-coupling superconductor not in the extreme dirty limit (we let $\ell\geq0.1\xi_0$), and system size $L$ small enough that $\tau_d\ll\tau_r$, $\tau_s$ holds and neglect electron-phonon processes.

\subsection{Quasiclassical theory}

To study stationary current flow in normal metal-superconducting heterostructures under voltage bias, we utilize the nonequilibrium quasiclassical Green's function method \cite{Eilenberger1968Apr,Larkin1969,belzig_quasiclassical_1999,eschrig_distribution_2000,Eschrig2009Oct}. All observables are then expressed in terms of three quasiclassical propagators. The retarded and advanced Green's functions contain information about the spectral properties, while the Keldysh Green's function also contains information about the nonequilibrium distribution function. In superconducting systems, the self-consistency condition on the order parameter leads to a coupling of spectrum and occupation that complicates the physical interpretation. Nevertheless, within the quasiclassical formulation, the two may be disentangled at the end of the calculation. The retarded (R), advanced (A), and Keldysh (K) propagators depend on momentum direction $\pF$ on the Fermi surface, coordinate $\RR$, and energy $\varepsilon$. We arrange them into a matrix in Keldysh space, denoted by a check,
\begin{equation}
\check{g}(\pF,\RR,\varepsilon) =
\begin{pmatrix}
\hat{g}^\mathrm{R}(\pF,\RR,\varepsilon) & \hat{g}^\mathrm{K}(\pF,\RR,\varepsilon)
\\
0 & \hat{g}^\mathrm{A}(\pF,\RR,\varepsilon)
\end{pmatrix}.
\end{equation}
Each propagator is in turn a matrix in combined particle-hole (Nambu) and spin spaces. We mark Nambu matrices by a hat ($\hat{~}$) and denote the three Pauli matrices in Nambu space as $\hat\tau_i$ and in spin space as $\sigma_i$, where in both cases $i=1,\,2\,$ and $3$.

The charge-current density is obtained from the Keldysh propagator:
\begin{align}
\mathbf{j}(\RR) &= e\NF\!\int\limits_{-\infty}^{\infty}\!
\frac{\mathrm{d}\varepsilon}{8\pi i} 
\FS{\mathrm{Tr}\left[ \vF \hat{\tau}_3 \hat{g}^\mathrm{K}(\pF,\RR,\varepsilon) \right]}
\label{eq:ChargeCurrentDefinition}
\\
&= \int\limits_{-\infty}^{\infty}\!\mathrm{d}\varepsilon\, \mathbf{j}(\RR,\varepsilon),
\end{align}
where $e=-|e|$ is the electron charge, $\NF$ is the density of states at the Fermi level in the normal state, and $\vF$ is the Fermi velocity. The trace is over both Nambu- and spin spaces. In the second line we also introduced the spectral current density $\mathbf{j}(\RR,\varepsilon)$. In three dimensions with a spherical Fermi surface, as we will assume in this paper, the average is given by integration over polar and azimuthal angles,
\begin{equation}
\FS{A(\pF)} = \frac{1}{4\pi} \int\limits_{0}^{2\pi}\!\mathrm{d}\varphi_\mathrm{F}\!\int\limits_0^\pi\!\mathrm{d}\theta_\mathrm{F} \sin\left(\theta_\mathrm{F}\right) A\left(\varphi_\mathrm{F}, \theta_\mathrm{F}\right)
.\label{eq:FermiSurfaceAverage}
\end{equation}
For a more complicated Fermi surface, see, e.g., Ref.~\cite{wennerdal_breaking_2020}.

The local electro-chemical potential is determined by a requirement of local charge neutrality \cite{Eschrig2009Jul}. It has the form
\begin{equation}
\phi(\RR) = \frac{1}{2e} \int\limits_{-\infty}^{\infty} \frac{\mathrm{d}\varepsilon}{8 \pi i} \FS{\mathrm{Tr}~\hat{g}^\mathrm{K}(\pF,\RR,\varepsilon)}.
\label{eq:phi}
\end{equation}
In equilibrium this potential is zero, while under steady-state non-equilibrium conditions it signals charge imbalance induced by injection of quasiparticles from normal-metal reservoirs. To achieve current conservation, it is required to compute $\phi(\RR)$ self-consistently with the self-energies.

The Green's functions are obtained by solving the quasiclassical transport equation,
\begin{equation}
i \hbar \vF \cdot \nabla \check{g} + \left[ \varepsilon \hat{\tau}_3 \check{1} - \check{h}, \check{g} \right] = 0,
\label{eq:transportequation}
\end{equation}
in combination with the normalization condition,
\begin{equation}
\check{g}^2 = - \pi^2 \check{1}.
\label{eq:normalization}
\end{equation}
Above and in the following, when there is no cause of confusion, we drop the explicit reference to the dependences on $\pF$, $\RR$, and $\varepsilon$.

The matrix $\check{h}$ with self-energies appearing in Eq.~\eqref{eq:transportequation} has the follwing elements in Nambu space:
\begin{align}
\hat{h}^{\mathrm{R}, \mathrm{A}} = 
\begin{pmatrix}
\Sigma & \Delta\\
\tilde{\Delta} & \tilde{\Sigma}
\end{pmatrix}^{\mathrm{R}, \mathrm{A}},
~
\hat{h}^\mathrm{K}
=
\begin{pmatrix}
\Sigma & \Delta
\\
-\tilde{\Delta} & -\tilde{\Sigma}
\end{pmatrix}^\mathrm{K}.
\end{align}
Objects with a tilde are related to non-tilde objects through particle-hole conjugation:
\begin{align}
\tilde{A}(\pF,\RR,\varepsilon) = A^*(-\pF,\RR,-\varepsilon^*).
\label{eq:TildeSymmetry}
\end{align}
In this paper, we include self energies for the mean-field superconducting order parameter as well as scalar and spin-flip impurities. They add up according to
\begin{equation}
\check{h} = \check{h}_\mathrm{mf} + \check{h}_\mathrm{s} + \check{h}_\mathrm{sf}.
\end{equation}

The superconducting order parameter is obtained through the self-consistency equation
\begin{equation}
\Delta_0(\RR) = \frac{\lambda \NF}{16 \pi i} \int\limits_{-\varepsilon_\mathrm{c}}^{\varepsilon_\mathrm{c}}\!
\mathrm{d}\varepsilon
\FS{\mathrm{Tr}\left[ i \sigma_2 ( \hat{\tau}_1 - i \hat{\tau}_2 ) \hat{g}^\mathrm{K}(\pF,\RR,\varepsilon) \right]},
\label{eq:DeltaSelfConsistencyEq}
\end{equation}
where $\lambda<0$ is the coupling strength, and $\varepsilon_\mathrm{c}$ is an energy cut-off. The spin-singlet symmetry of the order parameter is guaranteed by including the factor $i\sigma_2$ in the trace. Using the standard trick, where the linearized gap equation is added and subtracted, we eliminate the coupling strength $\lambda$ and the high-energy cutoff $\epsilon_\mathrm{c}$ from the gap equation in favor of the clean-limit superconducting transition temperature $\tc$. In Nambu space we now have
\begin{equation}
\hat h^{\mathrm{R}/\mathrm{A}}_\mathrm{mf}(\RR) = \left[\hat\tau_1\Re\left\{\Delta_0(\RR)\right\} - \hat\tau_2\Im\left\{\Delta_0(\RR)\right\}\right] i\sigma_2,
\end{equation}
while the Keldysh part is zero, $\hat{h}^\mathrm{K}_\mathrm{mf}=0$. The complex-valued order parameter can be written in terms of its magnitude and phase, $\Delta_0(\RR)=|\Delta_0(\RR)|\exp(i\chi(\RR))$. In practice we choose to work in a gauge where the order parameter is real and the phase enters through the superfluid momentum, defined as
\begin{equation}
\mathbf{p}_s(\RR) = \frac{\hbar}{2}\nabla\chi(\RR).\label{eq:ps}
\end{equation}
We assume that the lateral dimensions of the heterostructure are small compared with the penetration depth. In this case the back-coupling of the electromagnetic gauge field can be neglected.

Scattering against scalar impurities are taken into account within the self-consistent Born approximation,
\begin{align}
\check{h}_\mathrm{s}(\RR,\varepsilon) = \frac{\hbar}{2\pi \tau}
\FS{\check{g}(\pF,\RR,\varepsilon)},
\end{align}
where the scattering time $\tau$ is related to the mean free path via $\mfp=v_\mathrm{F}\tau$. Additional spin-flip impurities are included in normal-metal regions through the self-energy
\begin{align}
\check{h}_\mathrm{sf}(\RR,\varepsilon) = \frac{\hbar}{2\pi \tau_\mathrm{sf} }
\FS{\hat{\tau}_3 \check{g}(\pF,\RR,\varepsilon) \hat{\tau}_3}.
\end{align}
The mean free path for spin-flip scattering $\mfp_\mathrm{sf}=v_\mathrm{F}\tau_\mathrm{sf}$ will always be large compared with the mean free path due to scalar impurities, $\mfp_\mathrm{sf}\gg\mfp$, and is included in order for the proximity effect to decay. In this way we restrict ourselves to setups with normal-metal leads in Fig.~\ref{fig:sketch}(c) where superconducting coherence has decayed at the places we connect our device to reservoirs.

In the numerical implementation we use the Riccati parametrization of the Green's function \cite{eschrig_distribution_2000}. The normalization condition in Eq.~\eqref{eq:normalization} is then automatically fulfilled. The components of the retarded and advanced Green's functions are written as
\begin{align}
\hat{g}^{\mathrm{R},\mathrm{A}} 
=
\mp 2\pi i \begin{pmatrix}
\mathcal{G} & \mathcal{F}
\\
-\tilde{ \mathcal{F} } & - \tilde{ \mathcal{G} }
\end{pmatrix}^{\mathrm{R},\mathrm{A}}
\pm i \pi \hat{\tau}_3,
\label{eq:NambuSpaceGFsDef}
\end{align}
where $\mathcal{G}(\pF,\RR,\varepsilon)$ and $\mathcal{F}(\pF,\RR,\varepsilon)$ are expressed through the coherence functions $\gamma(\pF,\RR,\varepsilon)$ and $\tilde\gamma(\pF,\RR,\varepsilon)$ according to
\begin{align}
\mathcal{G} &= \left( 1 - \gamma \tilde{\gamma} \right)^{-1}, 
\\
\mathcal{F} &= \mathcal{G} \gamma.
\label{eq:RicattiGFDefinition}
\end{align}
The coherence functions are local amplitudes for conversion between particlelike and holelike branches ($e\leftrightarrow h$) and therefore express superconducting coherence. These functions satisfy the following Riccati equations
\begin{align}
\left( i\hbar\vF\cdot\nabla + 2 \varepsilon \right) \gamma^{\mathrm{R},\mathrm{A}}
&= \bigl( \gamma \tilde{\Delta} \gamma + \Sigma \gamma - \gamma \tilde{\Sigma} - \Delta \bigr)^{\mathrm{R},\mathrm{A}}\!,
\label{eq:GammaEquation}
\\
\bigl( i\hbar\vF\cdot\nabla - 2 \varepsilon \bigr) \tilde{\gamma}^{\mathrm{R},\mathrm{A}} 
&= \bigl( \tilde{\gamma} \Delta \tilde{\gamma} + \tilde{\Sigma} \tilde{\gamma} - 
\tilde{\gamma} \Sigma - \tilde{\Delta} \bigr)^{\mathrm{R},\mathrm{A}}\!.
\label{eq:GammaTildeEquation}
\end{align}
The components of $\hat{g}^\mathrm{K}$ in particle-hole space are labeled as
\begin{equation}
\hat{g}^\mathrm{K} = - 2 \pi i 
\begin{pmatrix}
\mathcal{X} & \mathcal{Y}
\\
\tilde{\mathcal{Y}} & \tilde{\mathcal{X}}
\end{pmatrix}^\mathrm{K}.
\end{equation}
We use a parametrization of these elements in terms of distribution functions $x(\pF,\RR,\varepsilon)$ and $\tilde{x}(\pF,\RR,\varepsilon)$,
\begin{align}
\mathcal{X}^\mathrm{K} &= \mathcal{G}^\mathrm{R} \left(~ x - \gamma^\mathrm{R} \tilde{x} \tilde{\gamma}^\mathrm{A} ~\right) \mathcal{G}^\mathrm{A},
\label{Eq:XKeldyshDefinition}
\\
\mathcal{Y}^\mathrm{K} &= \mathcal{G}^\mathrm{R} \left(~ x \gamma^\mathrm{A} - \gamma^\mathrm{R} \tilde{x} ~\right) \tilde{\mathcal{G}}^\mathrm{A}.
\end{align}
%
The distribution function $x$ satisfies the following transport equation:
\begin{align}
&i \hbar \vF \cdot \nabla x - \left[ \gamma \tilde{\Delta} + \Sigma \right]^\mathrm{R} x - x \left[ \Delta \tilde{\gamma} - \Sigma \right]^\mathrm{A}
\nonumber
\\
&=
- \gamma^\mathrm{R} \tilde{\Sigma}^\mathrm{K} \tilde{\gamma}^\mathrm{A} + \Delta^\mathrm{K} \tilde{\gamma}^\mathrm{A} + \gamma^\mathrm{R} \tilde{\Delta}^\mathrm{K} - \Sigma^\mathrm{K}.
\label{eq:XEquation}
\end{align}
The corresponding equation for $\tilde x$ can be obtained through Eq.~\eqref{eq:TildeSymmetry}.
We note that the coherence functions are in general matrices in spin space, but for the singlet superconducting case we study here they are simply proportional to $i\sigma_2$. The distribution functions, on the other hand, are trivially proportional to a unit matrix in spin space.
Note that the spatial coordinate along the system main axis is denoted $z$, as in Fig.~\ref{fig:sketch}, and should not be confused with the distribution function $x(\pF,\RR,\varepsilon)$.


In Eqs.~\eqref{eq:GammaEquation}-\eqref{eq:GammaTildeEquation} and Eq.~\eqref{eq:XEquation} each velocity direction $\vF(\pF)$ defines a trajectory in real space. We obtain the coherence functions and the distribution functions by integrating the first-order differential equations starting from initial conditions at the reservoirs. The equations for $\gamma^\mathrm{R}$ and $x$ are stable to integrate in the direction of $\vF$, while the equations for $\tilde\gamma^\mathrm{R}$ and $\tilde{x}$ are stable to integrate in the opposite direction. The opposite holds for the advanced coherence functions. We use the operator method presented in Ref.~\cite{Eschrig2009Oct} for stepping on a spatial grid with piece-wise constant self-energies \cite{grein_inverse_2013}.

Internal boundaries, such as tunnel barriers, have to be included through boundary conditions. These boundary conditions are given as Eqs.~(32)-(35) in Ref.~\cite{eschrig_distribution_2000}. The key input is the interface scattering matrix, which is expressed in terms of transmission and reflection amplitudes for incoming electrons in the normal state:
\begin{equation}
S(\pF) = \left(
\begin{array}{cc}
\sqrt{R(\pF)} & i\sqrt{D(\pF)}  \\
i\sqrt{D(\pF)} & \sqrt{R(\pF)}
\end{array}
\right),
\end{equation}
where $R(\pF)+D(\pF)=1$. The angle between the incoming momentum direction $\hat{p}_\mathrm{F}$ and the interface normal $\hat{n}$ define a directional cosine as $\hat{p}_\mathrm{F}\cdot\hat{n}=\cos\theta_\mathrm{F}$. The angular dependence of the transmission function is taken as a tunnel cone with the form
\begin{equation}\label{eq:tunnelcone_definition}
D(\theta_F)=D_0\frac{e^{-\beta\sin^2\theta_\mathrm{F}}-e^{-\beta}}{1-e^{-\beta}}.
\end{equation}
The parameter $\beta$ tunes the width of the cone, while $D_0$ sets the transparency for perpendicular incidence. 

\subsection{Distribution functions} 

In order to simplify the numerics, as well as getting more insight into the physics, it is convenient to split the distribution function into a local-equilibrium part $x^\mathrm{le}(\pF,\RR,\varepsilon)$ and a so-called anomalous part $x^\mathrm{a}(\pF,\RR,\varepsilon)$ by writing
\begin{equation}
x = x^\mathrm{le} + x^\mathrm{a}.
\label{eq:DistributionSplitting}
\end{equation}
The local-equilibrium part is defined according to
\begin{equation}
x^\mathrm{le} = F_0 + \gamma^\mathrm{R} \tilde{F}_0 \tilde{\gamma}^\mathrm{A},
\label{eq:x_le}
\end{equation}
where the familiar equilibrium form of the distribution function is
\begin{equation}
F_0(\RR,\varepsilon) = \tanh \frac{\varepsilon - e\phi(\RR)}{2\kb T}.
\label{eq:Equil_distrib}
\end{equation}
The remaining anomalous part of the distribution function, $x^\mathrm{a}$, captures the essential non-equilibrium effects. The splitting of $x$ according to Eq.~\eqref{eq:DistributionSplitting} carries over to a corresponding splitting of the Keldysh part of the Green's function:
\begin{equation}
\hat{g}^\mathrm{K} = \hat{g}^\mathrm{le} + \hat{g}^\mathrm{a}.
\label{eq:gKeldyshSplitting}
\end{equation}
This naturally leads to a separation of all observables that involve $\hat{g}^\mathrm{K}$ into a local equilibrium part and an anomalous part.

When this splitting is used, only the anomalous part of the distribution has to be stepped along trajectories since $x^\mathrm{le}$ is entirely determined by local quantities. For the stationary state, the equation of motion for $x^\mathrm{a}$ is given by Eq.~\eqref{eq:XEquation} with the replacements \cite{Eschrig2009Oct}
\begin{align}
x & \rightarrow x^\mathrm{a},
\label{Eq:XReplacement}
\\
\Delta^\mathrm{K} &\rightarrow \Delta^\mathrm{K} + \Delta^\mathrm{R} \tilde{F}_0 + F_0 {\Delta}^\mathrm{A},
\\
\Sigma^\mathrm{K} &\rightarrow \Sigma^\mathrm{K} - \bigl( \Sigma^\mathrm{R} - \Sigma^\mathrm{A} \bigr) F_0 + i \vF \nabla F_0.
\label{Eq:SigmaReplacement}
\end{align}

The distribution function $x(\pF,\RR,\varepsilon)$ can be related to the distribution function $h(\pF,\RR,\varepsilon)$ commonly used in literature \cite{Schmid1975Jul,Eschrig2009Oct}. The two are related through
\begin{align}
x = h + \gamma^\mathrm{R} \tilde{h} \tilde{\gamma}^\mathrm{A},
\end{align}
which has the inverse
\begin{align}
h = \sum\limits_{n = 0}^\infty \left( \gamma^\mathrm{R} \tilde{\gamma}^\mathrm{R} \right)^n \left( x - \gamma^\mathrm{R} \tilde{x} \tilde{\gamma}^\mathrm{A} \right) \left( \gamma^\mathrm{A} \tilde{\gamma}^\mathrm{A} \right)^n.
\end{align}
In a normal metal with $\gamma = 0$, the two distributions are identical, while in the presence of superconducting coherence this is no longer the case. The distribution $h$ can be divided into two parts with different structure in electron-hole space:
\begin{align}
\hat{f} =
\begin{pmatrix}
h & 0 
\\
0 & -\tilde{h}
\end{pmatrix}
= f_1 \hat{1} + f_3 \hat{\tau}_3.
\label{eq:h-splitting}
\end{align}
One obtains an electron distribution function through
\begin{align}
\fe = \tfrac{1}{2}(1 - h) = \tfrac{1}{2}\left(1 - f_1 - f_3\right),
\end{align}
which in equilibrium reduces to the Fermi-Dirac distribution function. The distribution $x$ can not be split in a similar fashion because it mixes particle and hole distributions. It is convenient to focus on $x$ and $\tilde{x}$ in the numerical implementation, while for interpreting the final results it may be beneficial to transform to the distribution $h$, or $f_{\mathrm{e}/1/3}$. In terms of $h$, the Keldysh Green's function $\hat{g}^\mathrm{K}$ has the form
\begin{align}
\hat{g}^\mathrm{K} = \hat{g}^\mathrm{R} \hat{f}  - \hat{f} \hat{g}^\mathrm{A}.
\label{eq:gKeldyshInH}
\end{align}
Correspondingly, all observables can be expressed either in terms of $x$ or $h$ types of distributions. This also makes the role of the two components $f_1$ and $f_3$ more transparent. The identities relevant for this paper are
\begin{equation}
\frac{\Delta_0(\RR)}{\lambda \NF/2}\! =\! \!\int\limits_{-\varepsilon_\mathrm{c}}^{\varepsilon_\mathrm{c}}\!\mathrm{d}\varepsilon
\FS{ f_1( \mathcal{F}^R + \mathcal{F}^A) - f_3( \mathcal{F}^R - \mathcal{F}^A) },
\label{eq:DeltaSelfConsistencyEq_hmatrix}
\end{equation}
for Eq.~\eqref{eq:DeltaSelfConsistencyEq},
\begin{equation}
\phi(\RR) = -\frac{1}{2e} \int\limits_{-\infty}^{\infty} \mathrm{d}\varepsilon \FS{f_3(\pF,\RR,\varepsilon) \mathcal{N}(\pF,\RR,\varepsilon)},
\label{eq:phi_hmatrix}
\end{equation}
for Eq.~\eqref{eq:phi}, and
%
\begin{align}
\mathbf{j}(\RR) &= -e\NF \int\limits_{-\infty}^{\infty}
\mathrm{d}\varepsilon
\FS{ \vF f_1(\pF,\RR,\varepsilon) \mathcal{N}(\pF,\RR,\varepsilon) },
\label{eq:ChargeCurrentDefinition_hmatrix}
\end{align}
for Eq.~\eqref{eq:ChargeCurrentDefinition}.
For the current we use as natural unit $j_0=e v_\mathrm{F}\NF\kb\tc$.
Above, $\mathcal{N}(\pF,\RR,\varepsilon)$ is a normalized momentum-resolved local density of states per spin:
\begin{equation}
\mathcal{N}(\pF, \RR, \varepsilon) = -\frac{1}{4\pi}\mathrm{Im} \mathrm{Tr}\left[\hat{\tau}_3 \hat{g}^R(\pF, \RR, \varepsilon)\right].
\end{equation}
The total local density of states is obtained as
\begin{equation}
N(\RR,\varepsilon) = 2\NF\FS{\mathcal{N}(\pF, \RR, \varepsilon)}.
\end{equation}
%
%

Finally, it is important to remember that all non-equilibrium distribution functions discussed above depend on momentum direction $\pF$. When impurity scattering is introduced, they become increasingly isotropic as the mean free path is reduced. It is then beneficial to study the Fermi surface averaged distributions:
\begin{align}
\fL(\RR,\varepsilon) &= \FS{f_1(\pF,\RR,\varepsilon)},\\
\fT(\RR,\varepsilon) &= \FS{f_3(\pF,\RR,\varepsilon)}.
\end{align}
They also have a certain symmetry with respect to energy. Following the nomenclature in Ref.~\cite{Schmid1975Jul}, $\fL$ is the longitudinal mode that is an odd function of energy, while $\fT$ is the transverse mode that is an even function of energy. In the diffusive limit, as in the Usadel formulation, these isotropic distributions also appear in observables and are the natural objects to study. In our formulation valid for arbitrary mean free path, only momentum resolved distributions, such as $x$, $h$, or $f_{\mathrm{e}/1/3}$, are used in the calculations. Their properties are different than $\fL$ and $\fT$ appearing in the diffusive Usadel theory. For instance, their symmetries are governed by Eq.~\eqref{eq:TildeSymmetry}.

\subsection{Local chemical potentials}\label{subsect:Noneq}

The local chemical potential $\phi(\RR)$ is given by Eq.~\eqref{eq:phi}. The local value $\phi(\RR)$ determines the local-equilibrium component $x^\mathrm{le}(\pF,\RR,\varepsilon)$ in Eq.~\eqref{eq:x_le}, and the driving terms for the anomalous component $x^\mathrm{a}(\pF,\RR,\varepsilon)$ in Eqs.~\eqref{Eq:XReplacement}-\eqref{Eq:SigmaReplacement}. We further define local right-mover $\phi_+(\RR)$ and left-mover $\phi_-(\RR)$ quasi-potentials as
\begin{equation}
\phi_{\pm} - \phi := -\frac{1}{2e}\int\limits_{-\infty}^{\infty}\! \frac{d\varepsilon}{4}~ \mathrm{Tr}_{\mathrm{spin}} \bigl[ \langle\mathcal{X}^\mathrm{a} \rangle_{\pm}  + \langle \tilde{\mathcal{X} }^\mathrm{a} \rangle_\mp \bigr], 
\label{eq:PhiLeftRight}
\end{equation}
where the label $+$ ($-$) indicates positive (negative) momentum projection along the $z$-axis and $\mathcal{X}^\mathrm{a}$ indicates that only the anomalous distribution $x^\mathrm{a}$ is used in Eq.~\eqref{Eq:XKeldyshDefinition}. The averages in \Eq{eq:PhiLeftRight} are defined as
\begin{align}
\langle A \rangle_\pm &\equiv \int\limits_{0}^{2\pi}\!\frac{d\varphi_\mathrm{F}}{2\pi}\!\int\limits_0^{1}\!d\xi_\mathrm{F} ~ A(\pm \xi_\mathrm{F}) ,
\label{eq:PositiveFermiAverage}
\end{align}
where $\xi_\mathrm{F}=\cos\theta_\mathrm{F}$. 

The two quasipotentials $\phi_\pm(\RR)$ in Eq.~\eqref{eq:PhiLeftRight} are measures of occupation of right- and leftmoving quasiparticle states relative to the local chemical potential $\phi(\RR)$. For systems not in the diffusive (Usadel) limit, these two quantities are more useful to develop an understanding of the transport behavior than the averaged chemical potential $\phi$, see also the discussion in Ref.~\cite{Datta2017Feb}. These potentials do not imply that the non-equilibrium distribution for left- and right-movers are simply Fermi distributions centered around these quasipotentials.
Note also that the averages over half the Fermi surface in Eqs.~\eqref{eq:PositiveFermiAverage} are missing a factor of $1/2$ in comparison to Eq.~\eqref{eq:FermiSurfaceAverage}. With this definition, in a normal metal
\begin{align}
\phi = \tfrac{1}{2}(\phi_+ + \phi_-),
\hspace{0.25cm}\mbox{(normal state)}
\label{eq:phi_av}
\end{align}
meaning that the measurable chemical potential $\phi$ is the average of left- and right-mover quasipotentials \cite{Datta2017Feb}. Note however, that Eq.~\eqref{eq:phi_av} is not satisfied in the superconducting state because of the proximity and inverse proximity effects.

\subsection{Voltage boundary condition}

In the case of voltage bias, we treat the normal regions as reservoirs at a chemical potential $\mu_\mathrm{L} = eV/2$ and $\mu_\mathrm{R} = -eV/2$, respectively. They are assumed to be unaffected by the contact to the central superconductor of length $L$. Thus, we are neglecting both the proximity effect as well as the effect of current flow from or into the normal leads.
Assuming such reservoirs, the distribution functions have the boundary conditions
\begin{align}
x(p_{\mathrm{F}z}>0,z=0,\varepsilon) = \tanh \frac{\varepsilon - eV/2}{2\kb T},\label{eq:distribL}\\
x(p_{\mathrm{F}z}<0,z=L,\varepsilon) = \tanh \frac{\varepsilon + eV/2}{2\kb T}.\label{eq:distribR}
\end{align}
The coherence functions incoming from the normal metal are assumed to vanish. Within the quantities introduced above, these distribution functions enforce a certain chemical potential of right-movers (left-movers) at the left (right) edge of the system.

\subsection{Current boundary condition}\label{sec:currentBC}

In left-right asymmetric systems, for instance an ISI structure with different interface barrier transparencies, or an NISIN structure with different lengths of the normal metals, we can not a priori know the potential profile. In this case we enforce a certain current $\Ib$ through the system boundaries and compute the resulting chemical potentials in left ($\mu_\mathrm{L}$) and right ($\mu_\mathrm{R}$) reservoirs, as well as the potential drop through the system self-consistently by requiring current conservation. The resulting voltage drop is $eV=\mu_\mathrm{L}-\mu_\mathrm{R}$, and would experimetally correspond to the externally applied voltage that result in the current $\Ib$.
We illustrate the procedure for the left system edge, where we need a boundary condition for the incoming distribution function $x_+$. 

To simplify the explanation, we assume a single-trajectory model so that only the two trajectories with $\xi_\mathrm{F} = \pm 1$ enter the calculation. The generalization to a full Fermi-surface average only adds trivial prefactors. Assuming that the normal-metal region is large enough for the proximity effect to decay, the current at the system edge can be written as
\begin{equation}
I = -\frac{e v_\mathrm{F} \Ac \NF}{2} \int\limits_{-\infty}^\infty\!d\varepsilon \left( x_+ - x_- \right) \equiv I_+ + I_-
\label{Eq:CurrentBoundaryDerivation}
\end{equation}
where we removed $\tilde{x}$ by symmetry. A possible choice for $x_+$ is
\begin{align}
x_+(z=0, \varepsilon) = \tanh \frac{\varepsilon - e \phi_+^\mathrm{b}}{2T},
\end{align}
where $\phi_+^\mathrm{b}$ is the boundary value that will be specified in the following. Using this choice and the requirement $I \stackrel{!}{=} \Ib$, we rewrite Eq.~\eqref{Eq:CurrentBoundaryDerivation} as
\begin{align}
\frac{\Ib - I_-}{-e v_\mathrm{F} \Ac \NF} = \frac{1}{2} \int\limits_{-\infty}^\infty\! d\varepsilon \tanh \frac{\varepsilon - e \phi_+^\mathrm{b}}{2T} = - e \phi_+^\mathrm{b}.
\label{eq:CurrentBoundaryFinal}
\end{align}
This can be used to determine $\phi_+^\mathrm{b}$ and thus the incoming distribution function $x_+(0)$ iteratively starting from the initial guess $x_-(0) = 0$. In each iteration we obtain a new value of $x_-(0)$ that adjusts $\phi_+^\mathrm{b}$ so that the total current at the edge is equal to $\Ib$. At the right system edge, the roles of $x_+$ and $x_-$ are swapped.

Once a self-consistent solution is obtained, we have charge-current conservation $I(z) = \Ib$ everywhere in the structure. The procedure determines the chemical potential at the left edge $\mu_\mathrm{L}=\phi_+^\mathrm{b}$ and right edge $\mu_\mathrm{R}=\phi_-^\mathrm{b}$ requiring a specified current $\Ib$ throughout the structure. In a completely normal-metal system the difference between the edge potentials will, for a given current, be dependent on the total resistance between the two ends. In the case of a central superconducting region, the respective edge potential will similarly be dependent on the resistance of the normal metal regions and interface resistances before the current becomes transformed into supercurrent in the superconductor, i.e., a finite $p_s(z)$.

This approach is particularly useful for asymmetric structures, where the bias drop over the structure does not have to be symmetric. This scheme is then a theoretical description of a four-point measurement: for a certain current flowing through the structure from source to drain leads, we determine the potential drop that would be measured by an additional high-resistance voltage probe. This scheme is thus highly relevant to experiments where a superconductor is dc-current biased and the potential drop is measured.

\subsection{Details on numerical procedure}\label{sec:numerics}

Performing a fully self-consistent nonequilibrium calculation is numerically challenging since Green's functions and self-energies have to be determined on the real axis. Here, we wish to outline some details on our numerical approach.

Firstly, all energies obtain a small imaginary part $\varepsilon \rightarrow \varepsilon \pm i \eta,$
when calculating retarded ($+$) and advanced ($-$) quantities. We chose $\eta = 10^{-3} \kb \tc$. In principle, this determines the required energy resolution to be $\Delta E \approx \eta$, and energies up to a cutoff energy $E_c \gg \Delta_0, \Gamma, e\phi$ have to be included. For the main results of our paper, we chose $E_c = 500 \kb \tc$. In order to reduce the resulting large number of energy points, we use a non-uniform grid that is more dense in an energy interval $\varepsilon \in (-5\Delta, 5 \Delta)$. Most of the physically relevant information is contained in this energy window. For example, the distribution function $x^a$ is only nonzero in the bias window $|\varepsilon| \leq \mu_\mathrm{L,R}$.


The number of iterations necessary to achieve selfconsistency is greatly reduced by iterating $p_\mathrm{s}(z)$, keeping the order parameter real, instead of iterating a complex-valued order parameter. The real order parameter is updated through the gap equation in Eq.~\eqref{eq:DeltaSelfConsistencyEq}. For the quasi-1D setup, we update the superfluid momentum $p_s(z)$ using the condition of current conservation:
\begin{align}
j(z) = \mathrm{const.} 
\label{eq:CurrentConservationCondition}
\end{align}
In the case of the current boundary condition discussed in Sect.~\ref{sec:currentBC}, the constant on the right-hand-side is the enforced boundary current $j_\mathrm{b}$. On the other hand, if a potential boundary condition is used then the right-hand side will be given by the current $j_\mathrm{int}$ at the interface between a normal and a superconducting region. Exactly at the interface grid points, both $\Delta_0$ and $p_\mathrm{s}$ have to be kept zero \cite{grein_boundary_2013}. On the normal side we specify both distribution and coherence functions while on the superconducting side the incoming functions are the result of propagating both functions through the self-energy landscape. 
An equivalent formulation of Eq.~\eqref{eq:CurrentConservationCondition} for the potential boundary condition is then
\begin{align}
\delta j(z) \equiv j(z) - j_\mathrm{int} = 0, 
\end{align}
with an analogous expression for the current boundary condition.
The local current deviation before selfconsistency, $\delta j(z)\neq 0$, can then be used to update $p_\mathrm{s}(z)$ throughout the superconductor. Starting from an initial guess $p_\mathrm{s} = 0$ everywhere in the superconductor, we use
\begin{align}
p_\mathrm{s}^{(n+1)} (z) = p_\mathrm{s}^{(n)} (z) + q ~ \frac{\kb\tc ~\delta j^{(n)} (z)}{v_\mathrm{F}j_0},
\end{align}
where $n$ is the iteration index and $q$ is the update step size of order unity. The interface current $j_\mathrm{int}$ will typically change during iterations but will eventually reach a fixpoint. In the results we present here, current is conserved up to a local relative error of $\delta j(z)/j_\mathrm{int} < 5\cdot 10^{-3}$ but higher accuracy can be achieved. 

In a majority of our calculations the Fermi surface average does not alter our results. We therefore use a one-trajectory model for simplicity, i.e., we keep $\theta_F=0$ and $\pi$ only. In relation to the tunnel cone in Eq.~\eqref{eq:tunnelcone_definition}, this implies $D=D_0$. In the few cases where the full angular dependence affects our results, we have chosen a wide tunnel cone with $\beta=1$.

\subsection{Example: Normal metal}\label{sec:normal}

Let us start with text book examples \cite{imry_introduction_2008,Datta2017Feb} in order to set our formalism into perspective. In a device as in Fig.~\ref{fig:sketch}(b) but consisting of normal-metal regions only, all superconducting coherence functions $\gamma^\mathrm{R}$ vanish and transport is described by the distribution functions only. The transport equation for $x$ simplifies to the Boltzmann equation for a normal metal,
\begin{equation}
i \hbar \vec{v}_F \cdot \vec{\nabla} x - (\Sigma^\mathrm{R}-\Sigma^\mathrm{A})x = - \Sigma^\mathrm{K}.\label{eq:Boltzmann}
\end{equation}
Assuming homogeneous current flow in the transverse direction, the current flowing along the wire main axis, quantified by a coordinate $z$, simplifies to 
\begin{align}
I &= -e\NF\Ac \int\limits_{-\infty}^\infty\!\frac{d\varepsilon}{2}
\FS{\vF ( x - \tilde{x})}\\
&= -e\NF\Ac \int\limits_{-\infty}^\infty\!d\varepsilon
\FS{\vF f_1},
\label{eq:Nstate_current}
\end{align}
where $\Ac$ is the wire cross sectional area. When only elastic impurity scattering is considered, combining Eqs.~(\ref{eq:Boltzmann})-(\ref{eq:Nstate_current}) implies conservation of spectral current, $j(\varepsilon) = -e\NF\FS{\vF f_1}$.
%
%
From these equations it is straight forward to perform the standard linear response calculation to an applied electric field along the $z$-axis, $E=-\partial_z\phi(z)$, with the local equilibrium distribution in Eq.~\eqref{eq:Equil_distrib}. The resulting charge conductivity is $\sigma_N = 2e^2\NF D$ for a dirty normal metal.

\begin{figure}[t]
    \centering
    \includegraphics{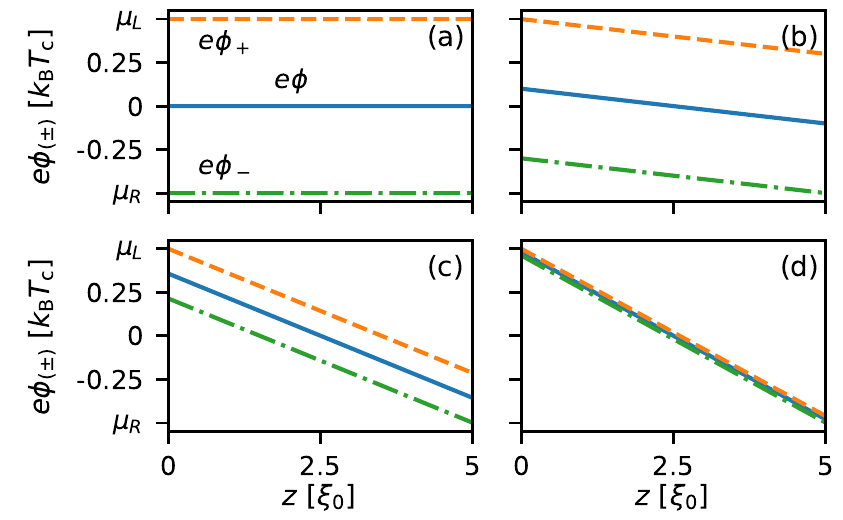}
    \caption{Spatial dependencies of the chemical potential $\phi$ (solid blue line) and quasi-potentials for right-movers $\phi_+$ (orange dashed line) and left-movers $\phi_-$ (green dash-dotted line). The applied bias is $eV/\tc = 1$, and the mean free path $l/\xi_0$ is (a) $\infty$, i.e., ballistic limit, (b) $10$, (c) $1$, and (d) $0.1$, diffusive limit.}
    \label{fig:PotentialDrop_N}
\end{figure}

For the device geometries, we go beyond linear response and perform calculations using assumptions of scattering theory with the device coupled to reservoirs \cite{imry_introduction_2008}. In Fig.~\ref{fig:PotentialDrop_N} we show how the chemical potential, as well as left- and right-mover quasi-potentials, drop across a normal metal piece between two reservoirs placed at $z=0$ and $z=5\xi_0$, for different mean free paths. We note that the natural length scale is the mean free path, but although there is no superconductor here we use the coherence length to be consistent with later sections of the paper.

In the clean case ($l\rightarrow\infty$), see Fig.~\ref{fig:PotentialDrop_N}(a), the incoming left- and right-moving populations simply shoot through the structure and do not mix, hence $\phi_+=\mu_L$, $\phi_-=\mu_R$, and $\phi = 0$ everywhere. This corresponds to a Landauer-B\"uttiker wavefunction based scattering approach for the trivial case of no scattering in the device.

In the diffusive limit, in contrast, the chemical potential $\phi$ at the system edges equals the reservoir chemical potentials and in between interpolates linearly between the two. There is no difference between left- and right-movers quasi-potentials in the diffusive limit, see Fig.~\ref{fig:PotentialDrop_N}(d), since isotropization is to lowest order approximation local ($l\ll L$).
The fully diffusive limit can also be obtained from the Usadel equation \cite{Keizer2006Apr}. The solution for the isotropic distribution function reads
\begin{align}
x_\mathrm{diff}(z) = \frac{z}{L}(x_\mathrm{L} - x_\mathrm{R}) + x_\mathrm{L},
\end{align}
where $x_\mathrm{L}$ and $x_\mathrm{R}$ are the left-edge and right-edge boundary values, respectively. This leads to 
\begin{align}
e\phi_\mathrm{diff}(z) = \frac{z}{L}(e\phi_\mathrm{L} - e\phi_\mathrm{R}) + e\phi_\mathrm{L},
\end{align}
so that in the diffusive limit, the boundary condition for $\phi_\pm$ at the edges becomes effectively boundary conditions for $\phi$. The solution is then indeed in agreement with Fig.~\ref{fig:PotentialDrop_N}(d).

For mean free paths in between these two limiting cases, as in Fig.\ref{fig:PotentialDrop_N}(b) and Fig.~\ref{fig:PotentialDrop_N}(c), the behaviour of $\phi$ as well as $\phi_\pm$ interpolate between the ballistic and diffusive cases. We note that since we compute the chemical potential $\phi$ everywhere in the device, its edge values will disagree with the reservoir chemical potentials. It is well known \cite{imry_introduction_2008,Datta2017Feb} that this differences correspond to the contact potentials, or spreading resistance \cite{boogaard_resistance_2004}, via the adiabatic and ballistic leads to the reservoirs. Since we go beyond linear response in our treatment, we need to have well-defined reservoirs with well-defined incoming distribution functions ($\phi_\pm^b$), in agreement with the Landauer-B\"uttiker approach \cite{imry_introduction_2008}.

\subsubsection{Asymmetric system: current boundary condition}

\begin{figure}
    \centering
    \includegraphics{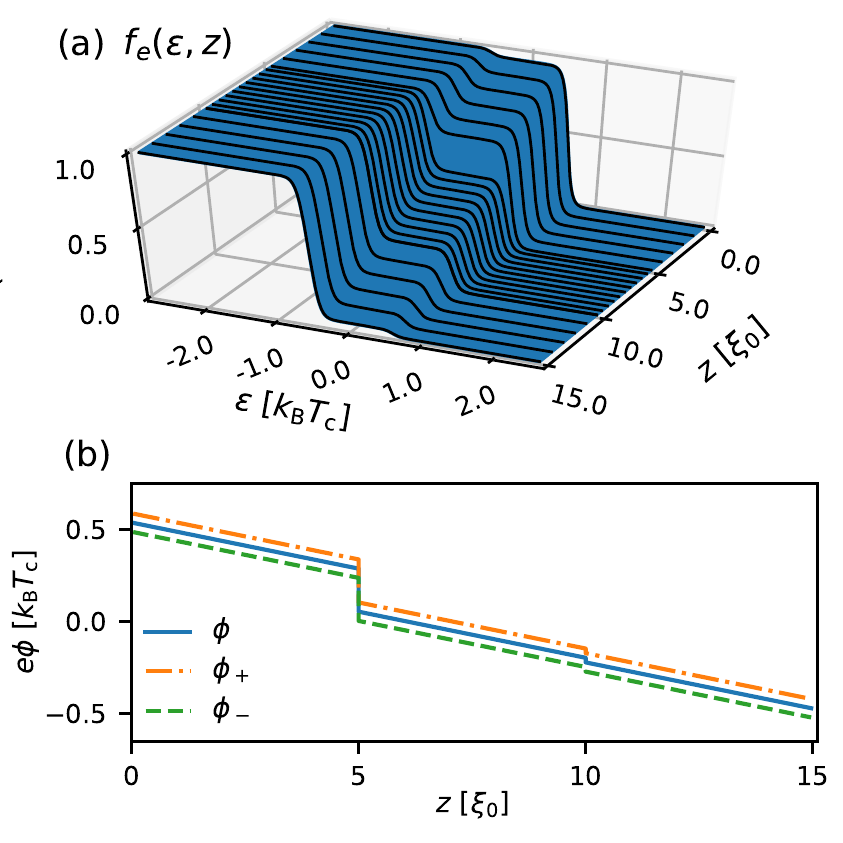}
    \vspace{-20px}
    \caption{(a) Distribution function for a boundary current of $I_\mathrm{b} = 0.1I_0$ across an asymmetric system with barrier transparencies $D_L=0.3$ and $D_R = 0.8$, and a mean free path of $\ell=\xi_0$. (b) The potential drop for the same set of parameters, the total voltage drop is $eV=\mu_L-\mu_R \approx 1.356 \kb \tc$.
    }
    \label{fig:distributionComp_CurrentBias}
\end{figure}

Next let us study an asymmetric device of total length $L=15\xi_0$, with two internal barriers at $z=5\xi_0$ and $z=10\xi_0$ of transparencies $D_\mathrm{L}=0.3$ and $D_\mathrm{R} = 0.8$, see Fig.~\ref{fig:distributionComp_CurrentBias}. In this case, since we a priori do not know how the voltage drops across the system, we need to enforce a current and let the chemical potentials at left and right edges float up and down to reach self-consistent values while enforcing current conservation across the system. An example of the resulting total voltage drop $eV$, chemical potential profile $\phi(z)$, and the corresponding electronic distribution function $\fe(\epsilon,z)$ are shown in Fig.~\ref{fig:distributionComp_CurrentBias} for an intermediate mean free path $l/\xi_0=1$ and an enforced current $I_\mathrm{b}=0.1I_0$, where $I_0=j_0\Ac$. This result for the distribution function agrees well with the experiment by Pothier {\it et al.} \cite{pothier_energy_1997}.

Having established how our formalism works in the normal state, we shall from now on focus on the case when the central region (in this example $z/\xi_0\in[5,10]$) is superconducting.

\section{Results}

\subsection{I-S system}\label{subsec:IS}

\begin{figure}[t]
    \centering
    \includegraphics[scale=1.0]{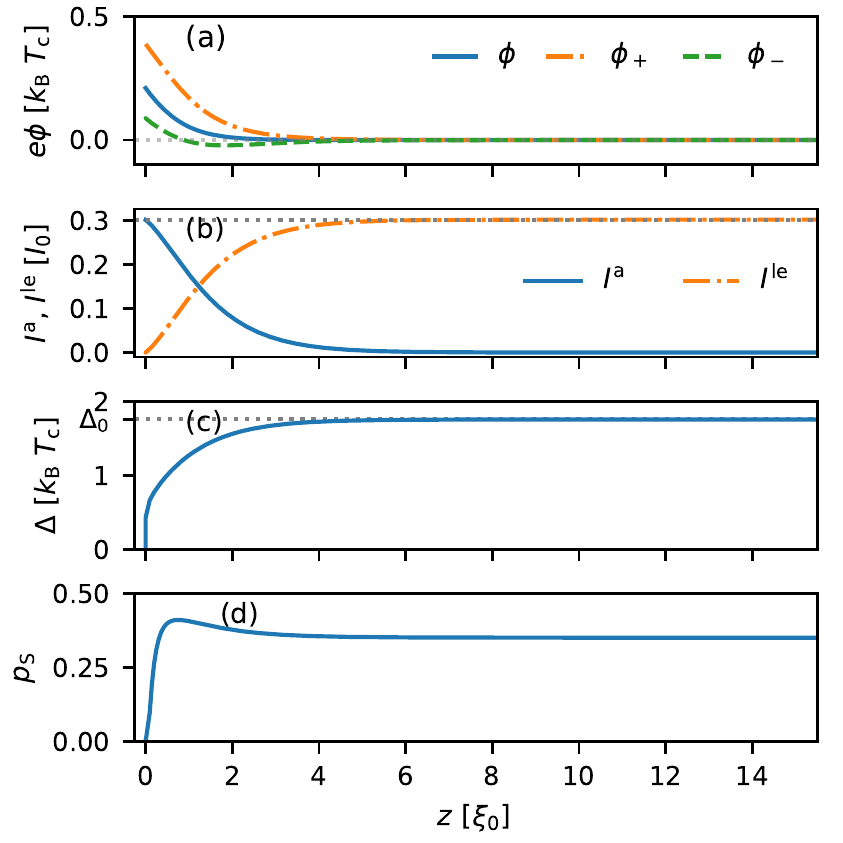}
    \caption{Spatial dependencies of the main physical quantities in an IS system for barrier transparency $D=0.8$, mean free path $\mfp = \xi_0$, temperature $T=0.01T_c$, and applied bias $eV= 0.5\kb\tc$. (a) Left-mover and right-mover quasi-potentials $\phi_\pm$ and chemical potential $\phi$. (b) Anomalous, or quasiparticle, current (solid line) and local-equilbrium, or condensate, current (dash-dotted line). (c) Order parameter $\Delta_0$. (d) Superfluid momentum $p_\mathrm{s}$.}
    \label{fig:observables_117_dirty}   
\end{figure}

The first setup in Fig.~\ref{fig:sketch}(a) consists of contacting the superconductor on one side to a normal-metal reservoir through a tunnel barrier. The applied voltage sets the chemical potential in the normal metal to $\mu_L=eV$ relative to the superconducting reservoir $\mu_R=0$. The contact specifies a boundary condition for the incoming distribution function $x_+(z=0^-,\varepsilon)$ according to Eq.~\eqref{eq:distribL}.

The boundary condition for the spectral part is $\gamma^R(z=0^-,\varepsilon)=0$, i.e., the proximity effect is neglected in the normal-metal side in agreement with the reservoir assumption. If self-consistency and current conservation is neglected in the superconductor we obtain the BTK result for the interface conductance \cite{Blonder1982Apr}. We go beyond this approximation and examine the effect of current conservation as well as impurity scattering on transport.

The first thing to note is that Anderson's theorem \cite{Anderson1959Sep}, which states that the superconducting order parameter is unaffected by scalar impurities, relies on time-reversal symmetry. Therefore it will not hold in the presence of current flow. Even far from the contact, the current flow induces Doppler shifts of right- and left-moving quasiparticles according to
\begin{equation}
\varepsilon \rightarrow \varepsilon - \vF\cdot\ps,
\label{eq:DopplerShift}
\end{equation}
where the superfluid momentum $\ps$ is set by the gradient of the order parameter phase $\chi$ as defined in Eq.~\eqref{eq:ps}. These Doppler shifts lead to violation of Anderson's theorem and the order parameter depends on the mean free path.

\begin{figure*}[t]
    \centering
    \includegraphics[scale=1.0]{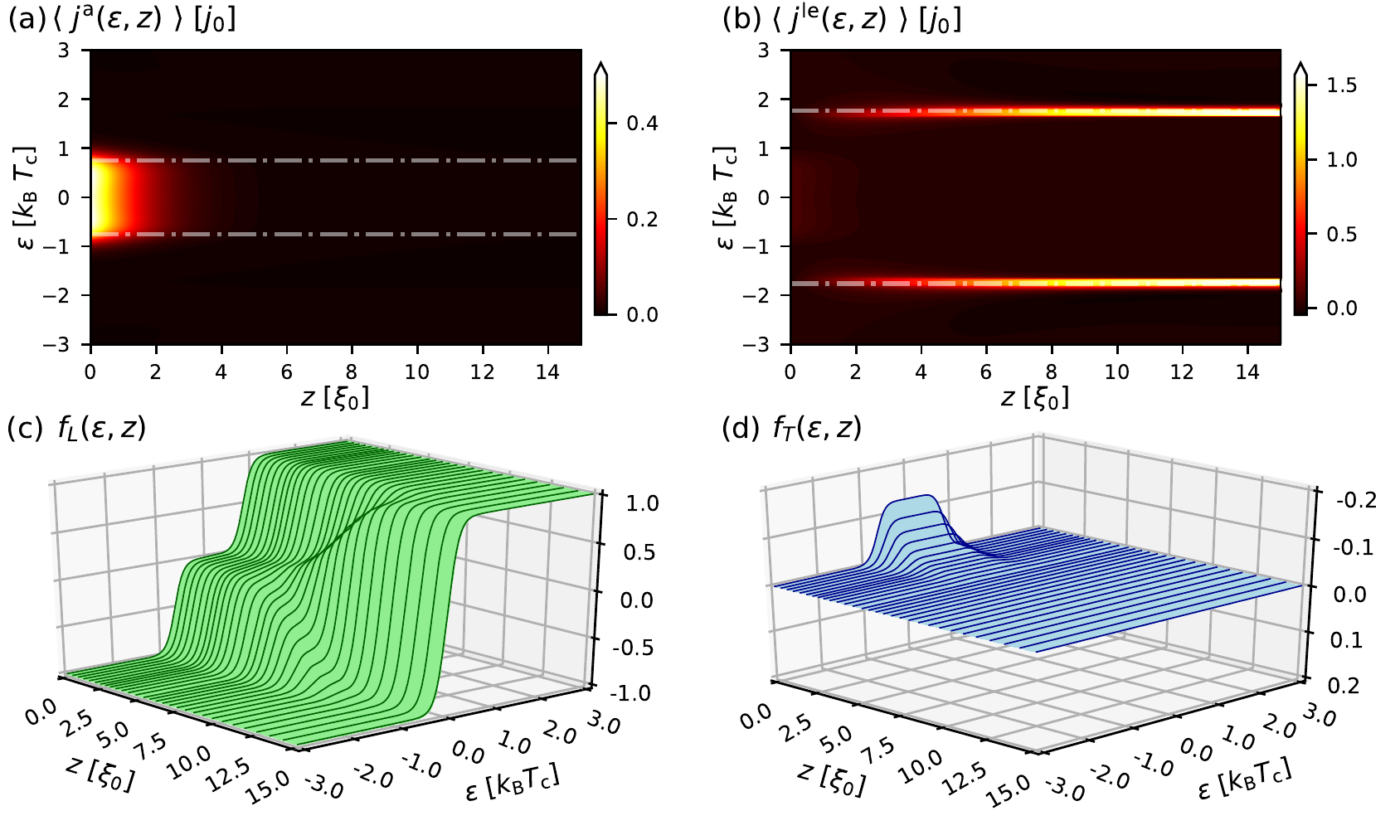}
    \caption{(a) Anomalous and (b) local-equilibrium spectral current density across the IS device.  (c) Transverse and (d) longitudinal modes of the distribution function near the IS interface on the superconducting side. In (a), the white dashed-dotted line indicated the width of the bias window $eV = 0.75 ~\kb \tc$, while the line in (b) indicates the width of the equilibrium bulk energy gap $\Delta_0$. All other parameters are the same as in Fig.~\ref{fig:observables_117_dirty}.}
    \label{fig:IS_distributions}
\end{figure*}

In Fig.~\ref{fig:observables_117_dirty} we show results of a self-consistent calculation for one particular set of parameters and for one voltage. The boundary condition for the distribution function enforces a certain anomalous current $\Ia$ at the interface that slowly decays into the superconductor, see Fig.~\ref{fig:observables_117_dirty}(b). Since current is conserved, this decay is compensated for by a corresponding increase in the local-equilibrium current $\Ile$. The conversion is associated with the appearance of the chemical potential $\phi(z)$ near the interface, see Fig.~\ref{fig:observables_117_dirty}(a). The local-equilibrium current is carried by a finite, position-dependent superflow in the superconductor, meaning that the order parameter phase varies with coordinate and the superfluid momentum is non-zero, see Fig.~\ref{fig:observables_117_dirty}(d). The suppression of the order parameter close to the interface, shown in Fig.~\ref{fig:observables_117_dirty}(c), leads to a peak in the superfluid momentum $\ps(z)$ before both reach their bulk values roughly $10\xi_0$ away from the interface. The absolute value of the total current depends on interface barrier transparency, impurity concentration in the superconductor, and the applied voltage.

The spectral current and the distribution functions are shown in Fig.~\ref{fig:IS_distributions}. The anomalous current in Fig.~\ref{fig:IS_distributions}(a) is due to the quasiparticle injection in the subgap region within a voltage window $2eV$. Through Andreev reflection this current is converted to superflow, which is carried by continuum states, see Fig.~\ref{fig:IS_distributions}(b). In the interface region, where the potential $\phi$ is non-zero, also the transverse distribution function $f_T$ is non-vanishing as shown in Fig.~\ref{fig:IS_distributions}(d). Well inside the superconductor it has decayed to zero. At the same time the longitudinal distribution decays back to the equilibrium form and the supercurrent is only due to the Doppler shifts in Eq.~\eqref{eq:DopplerShift} of continuum states. 

\begin{figure}[t]
    \centering
    \includegraphics[scale=1.0]{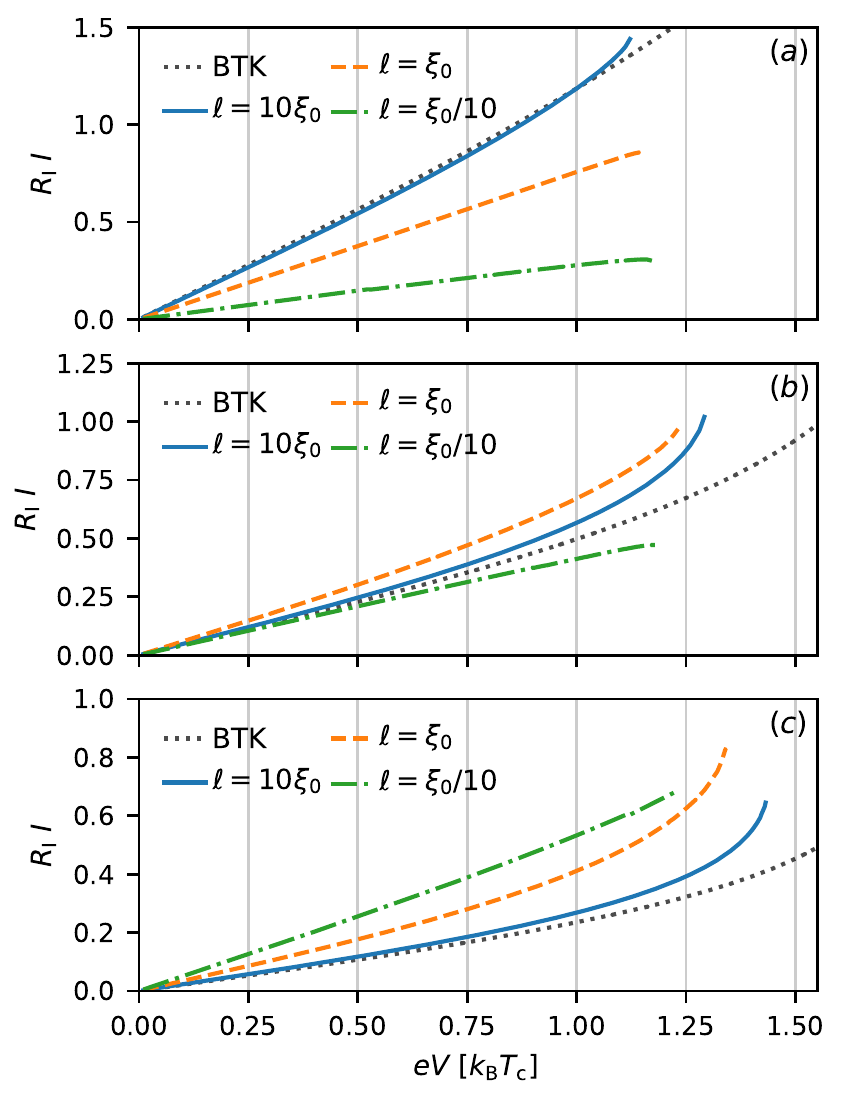}
    \caption{Current-voltage curves for the IS system with different mean free paths from clean to dirty limits and barrier transparencies (a) $D=0.8$, (b) $D=0.5$, and (c) $D=0.3$ in comparison to the BTK clean limit non-selfconsistent current formula (dotted lines).}
    \label{fig:CurrentCurves_dirty}
\end{figure}

In Fig.~\ref{fig:CurrentCurves_dirty} we display the total current as function of applied bias voltage for three different interface transparencies, and three different mean free paths. For the clean case (solid lines, $\ell=10\xi_0$) the current-voltage characteristics agree well with the BTK results (dotted lines), in particular at low voltage. Corrections appear at higher voltage and are mainly due to the finite Doppler shifts. The solid lines end where superconductivity near the interface starts to break down, an effect that we will study in more detail below.

In the case of a highly transmissive barrier, $D = 0.8$ in Fig.~\ref{fig:CurrentCurves_dirty}(a), the current gets reduced with shorter mean free path. In the presence of impurities, quasiparticles that are transmitted through the interface can be scattered back before they get Andreev-reflected. For a given injection energy $eV$, Andreev reflection happens on a lengthscale $\xi_\mathrm{AR} = \hbar v_\mathrm{F}/\sqrt{\Delta^2 - eV^2}$, while scattering happens on the scale of $\mfp$. The reduction of Andreev reflection and current is thus stronger for higher impurity concentration and at higher voltages.

In contrast, the current is increasing with reduced mean free path for barriers with lower transparency, see the cases of $D=0.5$ and $D=0.3$ in Fig.~\ref{fig:CurrentCurves_dirty}(b) and Fig.~\ref{fig:CurrentCurves_dirty}(c). This behavior can be understood from two points of view. First, by including impurity scattering the subgap local density of states becomes finite near the interface, even in equilibrium, due to the so-called inverse proximity effect that occurs on the coherence length scale \cite{belzig_local_1996,boogaard_resistance_2004}. From a transport perspective, Andreev reflection is already reduced by the low barrier transparency. On the other hand, normal reflection gets reduced by impurity scattering. The impurity scattering filling the subgap energies with states is associated with a finite normal electron transmission amplitude locally at the interface, which is not present in the ballistic case. This enables a larger current for the same applied voltage. Note that in the bulk superconducting region there is only superflow, meaning that the local finite amplitude of normal transmission at the interface is not associated with quasiparticle flow in the asymptotic region, which also remains fully gapped.

In the experiment presented in Ref.~\cite{boogaard_resistance_2004}, the resistance of an Al wire was measured as function of temperature. The fit to Usadel theory was good, but at low temperature the resistance was lower than what was predicted by theory. According to Ref.~\cite{boogaard_resistance_2004} the transparencies of the interfaces to the normal-metal reservoirs were very high, corresponding to our Fig.~\ref{fig:CurrentCurves_dirty}(a).
We predict that the correction to the Usadel result from ballistic effects is a decrease of the IS resistance [enhanced conductance in Fig.~\ref{fig:CurrentCurves_dirty}(a)]. This follows the trend seen in the experiment at low temperature.

\subsubsection{Critical voltage}

All of the above current-voltage curves end at certain voltage points that we refer to as critical voltages $\Vc$ following Ref.~\cite{Keizer2006Apr}. At the critical voltage the order parameter either vanishes throughout the system, i.e. superconductivity breaks down, or self-consistency can not be achieved. The latter case happens in the clean limit for high barrier transparency, where the current is sufficiently high that the Doppler shifts lead to injection into continuum states. In this case, the potential $\phi(z)$ also extends out into the bulk of the superconductor. We will return to this case in the next section, when we have two normal-metal reservoirs and a more well defined set-up.
\begin{figure}[t]
    \centering
    \includegraphics[scale=0.98]{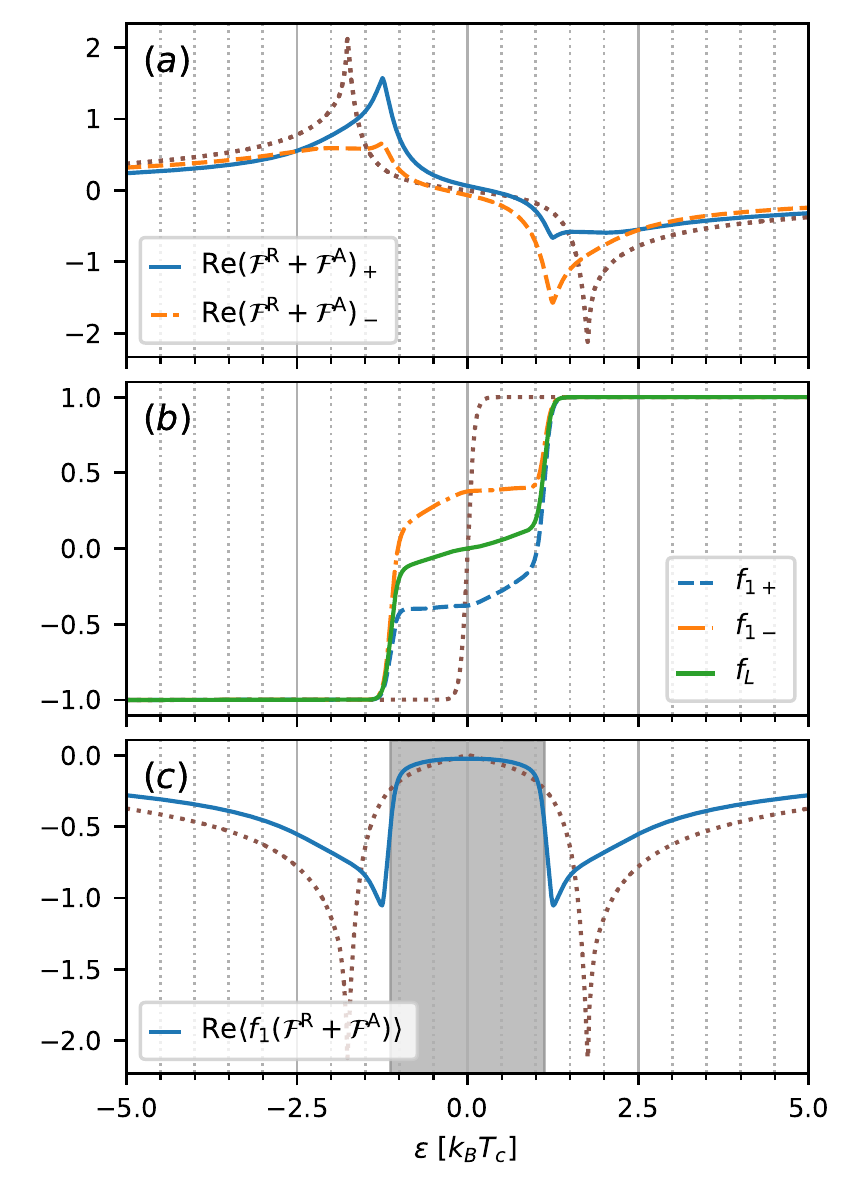}
    \caption{Real part of the constituting factors of the self-consistency integral kernel in Eq.~\eqref{eq:DeltaSelfConsistencyEq_hmatrix}. All functions are at a distance $z = 2\xi_0$ from the barrier with transparency $D=0.8$. The mean free path is $\mfp=\xi_0$ and the applied voltage is $eV = 1.13\tc$, which is close to the critical voltage for the dashed curve in Fig.~\ref{fig:CurrentCurves_dirty}(a).
    (a) Anomalous Green's functions $\mathrm{Re}\left( \mathcal{F}^R + \mathcal{F}^A \right)$ for left- and right-movers compared to their (identical) equilibrium values (red, dotted line).
    (b) Non-equilibrium energy-mode $\fL$ for left- and right-movers as well as their angular average. In equilibrium the three distributions coincide with the equilibrium distribution shown as a red, dotted line.
    (c) The product of the spectral parts and energy mode distributions, as entering the self-consistency equation. The equilibrium form is shown as the red, dotted line. The grey-shaded region indicates the bias-window in the non-equilibrium case.}
    \label{fig:DeltaKernelFigureL1}
\end{figure}

The breakdown behavior can be understood by a reexamination of the self-consistency equation for the  order parameter $\Delta_0$. Using the parametrization of $\hat{g}^K$ in terms of distribution functions $f_1$ and $f_3$ in Eq.~\eqref{eq:h-splitting}, the self-consistency equation for $\Delta_0$ can be written as in Eq.~\eqref{eq:DeltaSelfConsistencyEq_hmatrix}.

Both terms consist of a spectral contribution, $\mathcal{F}^R \pm \mathcal{F}^A$, and either the energy-like mode $f_1$ or the charge-like mode $f_3$. Note that both distributions are real functions. We use a gauge where $\Delta_0$ is real and obtain $\ps$, which gives the spectral rearrangements. For the real part of the kernel in the right hand side, we find that the first term gives the dominating contribution to $\Delta_0$. Figure~\ref{fig:DeltaKernelFigureL1} shows the real part of the first term of the kernel in Eq.~\eqref{eq:DeltaSelfConsistencyEq_hmatrix} for a voltage close to $\Vc$. For comparison, the equilibrium forms are included as dotted lines. In Fig.~\ref{fig:DeltaKernelFigureL1}(a) we show the factor $\mathrm{Re}\left(\mathcal{F}^R+\mathcal{F}^A\right)$, while in Fig.~\ref{fig:DeltaKernelFigureL1}(b) we display the distribution function $f_1$ for right- and left-moving states. The main effect of the current injection is that the distribution function is reduced in an energy window of width $2eV$ around zero energy. In addition, the coherence peaks in $\mathrm{Re}\left(\mathcal{F}^{R}+\mathcal{F}^A\right)$ are Doppler shifted towards lower energies. As a result, the total kernel gets reduced with increasing voltage, see Fig.~\ref{fig:DeltaKernelFigureL1}(c). For voltages approaching $\Vc$, the coherence peaks gets shifted inside the energy window where $f_1$ is strongly reduced. This leads to a suppression of the order parameter and eventually turns the system normal.

\begin{figure}[t]
    \centering
    \includegraphics[scale=1.0]{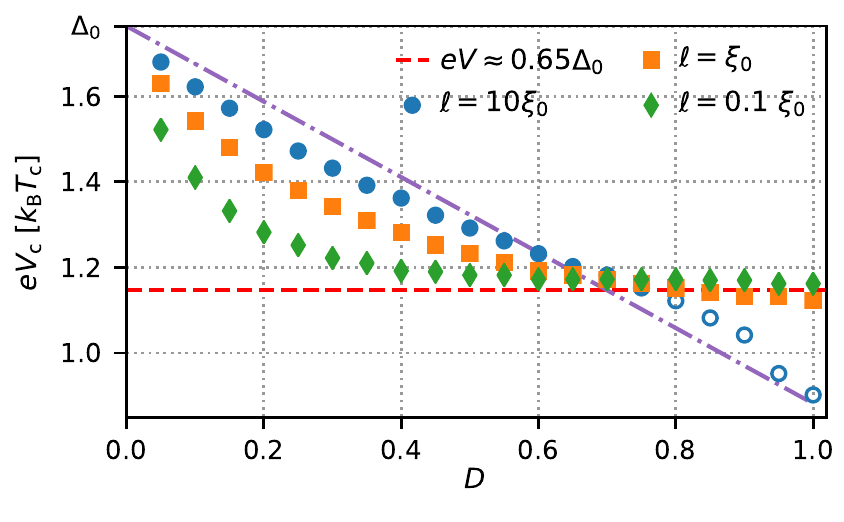}
    \caption{Critical voltage of the IS system as function of interface transparency for different impurity strengths. The dash-dotted line given by $eV_c = \Delta_0(1 - D/2)$ is a guide to the eye.}
    \label{fig:CriticalVoltage_dirty}
\end{figure}

We summarize in Fig.~\ref{fig:CriticalVoltage_dirty} the critical voltage as function of barrier transparency for three mean free paths from the clean to the dirty limit. The general trend is that lower transparency barriers lead to a higher critical voltage. This is due to the reduced current and reduced Doppler shifts for the same applied voltage at low transparency. For transparencies $D \gtrsim 0.75$, we observe an increase in the critical voltage with decreasing mean free path. The reduction in current with decreasing mean free path, shown in Fig.~\ref{fig:CurrentCurves_dirty}(a) above for $D=0.8$, means that $p_s$ goes down and the coherence peaks get shifted less, allowing for a larger critical voltage. In the ballistic case, a high-transparency interface can have steady-state solutions with non-decaying chemical potential in the bulk of the superconductor. In this case we inject into continuum states and the non-equilibrium population does not decay deep in the superconductor since we neglect inelastic relaxation processes. This situation for the IS set-up is beyond the scope of this paper, but we return to this case in the next section where we have two reservoirs and a well defined set-up. The corresponding points where injection into continuum state appear are marked by empty symbols rather than filled ones in Fig.~\ref{fig:CriticalVoltage_dirty}.

In contrast, the critical voltage decreases with impurity concentration for transparencies $D \lesssim 0.75$. We understand this result in the light of Fig.~\ref{fig:CurrentCurves_dirty}. For smaller transparencies ($D=0.5$ and $D = 0.3$) the current for a particular applied voltage increases for smaller mean free paths. This leads to more pair breaking and a lower critical voltage.

Finally, we note that in the diffusive limit, the critical voltage is maximal in the tunneling limit and decreases towards a minimum at full transparency. For the case of $D=1$, we recover the critical voltage for the diffusive NSN system in the zero-temperature limit \cite{Keizer2006Apr}, indicated by the red, dashed line.

\subsection{I-S-I system}\label{subsec:ISI}

\begin{figure}[t]
    \centering
    \includegraphics[scale=1.0]{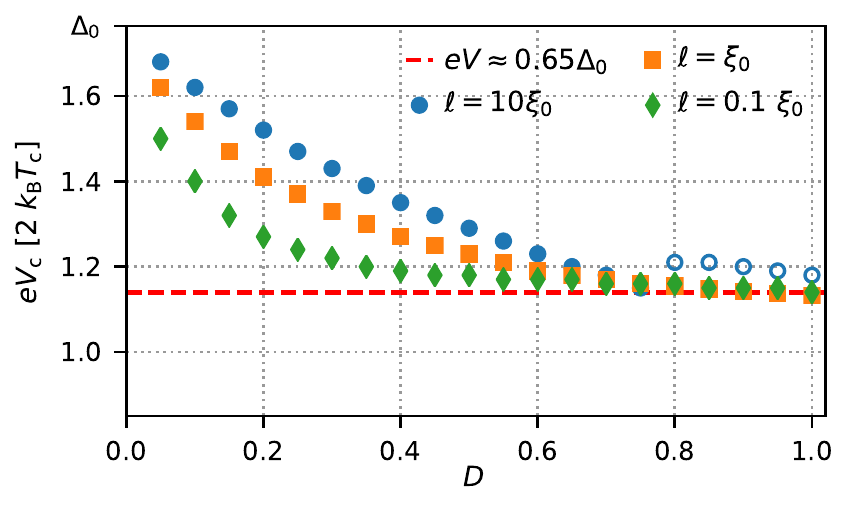}
    \caption{Critical voltage of the ISI system as function of interface transparency $D_\mathrm{L}=D_\mathrm{R}=D$ for different impurity strengths. The unfilled symbols indicate oscillating solutions before the transition to the normal state. The length of the S-region is $L=100\xi_0$ for mean free paths $\mfp=10\xi_0$ and $\mfp=\xi_0$, while it is $L=50\xi_0$ for $\mfp=0.1\xi_0$.}
    \label{fig:CriticalVoltage_dirty_ISI}
\end{figure}

\begin{figure}[t]
    \centering
    \includegraphics[scale=1.0]{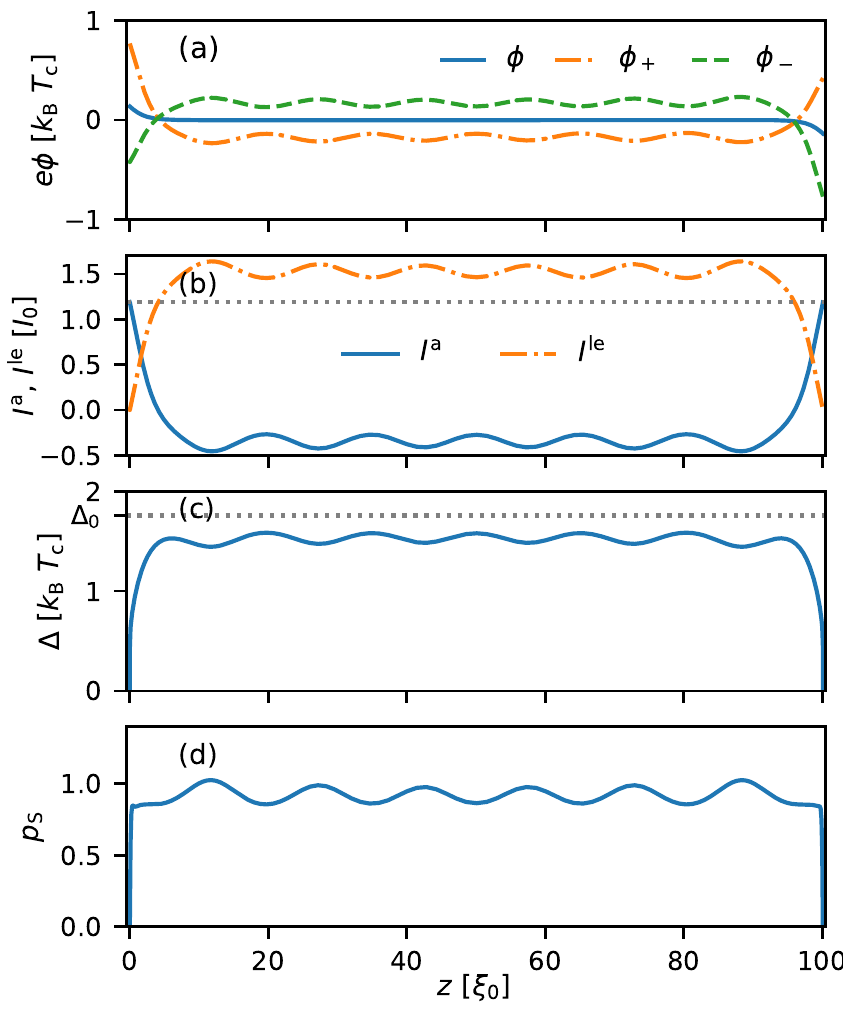}
    \caption{Spatial dependencies of the main physical quantities for the ISI system with barrier transparencies $D=0.8$, mean free path $\mfp = 10\xi_0$, and applied bias $eV = 2.3\kb\tc$, near the critical voltage. For this setup, electron-hole transmission through continuum states is possible, leading to interference and oscillations in all quantities. (a) Left-mover and right-mover quasi-potentials $\phi_\pm(z)$ and average chemical potential $\phi(z)$. (b) Anomalous and local-equilbrium currents, (c) order parameter $\Delta_0$(z), and (d) superfluid momentum $p_\mathrm{s}(z)$.}
    \label{fig:observables_115_clean}
\end{figure}

In the ISI case, we choose a symmetric system with $D_\mathrm{L} = D_\mathrm{R}=D$, and a symmetric bias of $\mu_\mathrm{L} = - \mu_\mathrm{R} = eV/2$. 
Similar to the IS system, we find a critical voltage where superconductivity breaks down which, in turn, depends on interface transparency and mean free path, see Fig.~\ref{fig:CriticalVoltage_dirty_ISI}. The main new result compared with the IS case is that for relatively clean systems with high-transparency barriers $D\gtrsim 0.8$, we can study solutions with non-vanishing $\phi$ everywhere in the superconductor. Such solutions typically have oscillations in both the order parameter, the superfluid momentum, and the division between anomalous and local-equilibrium currents, see Fig.~\ref{fig:observables_115_clean}. Such solutions occur when the Doppler shifts are so large that quasihole continuum states are shifted into the bias window. Under voltage bias these states become occupied through an electron-hole transmission process. Within the 1D model, these states interfere with returning quasielectron continuum states which results in oscillations analogous to Tomasch oscillations, with a voltage-dependent wavelength that does not depend on system size. For shorter mean free paths of the order of the oscillation period, the interference is suppressed. In the diffusive limit they are absent.

Going beyond the 1D model by performing a full circular Fermi-surface average, the oscillations survive for barriers with a wide tunnel cone, while they are suppressed for narrow ones. The narrow tunnel cone leads effectively to a lower barrier transparency and less dramatic Doppler shifts. The transition from non-oscillating to oscillating solutions becomes less sharp as function of voltage. The quasihole states are not available for all trajectories at once due to the angular dependency of the Doppler shift in Eq.~\eqref{eq:DopplerShift}. Typically, small-scale oscillations set in first,
and increase in magnitude as more trajectories enter the bias window with increasing applied voltage. In summary, only the quasi-1D model with a wide tunnel cone can display pronounced oscillations.

\subsection{N-I-S-I-N system}\label{subsec:NISIN}

\begin{figure}[t]
    \centering
    \includegraphics[scale=1.0]{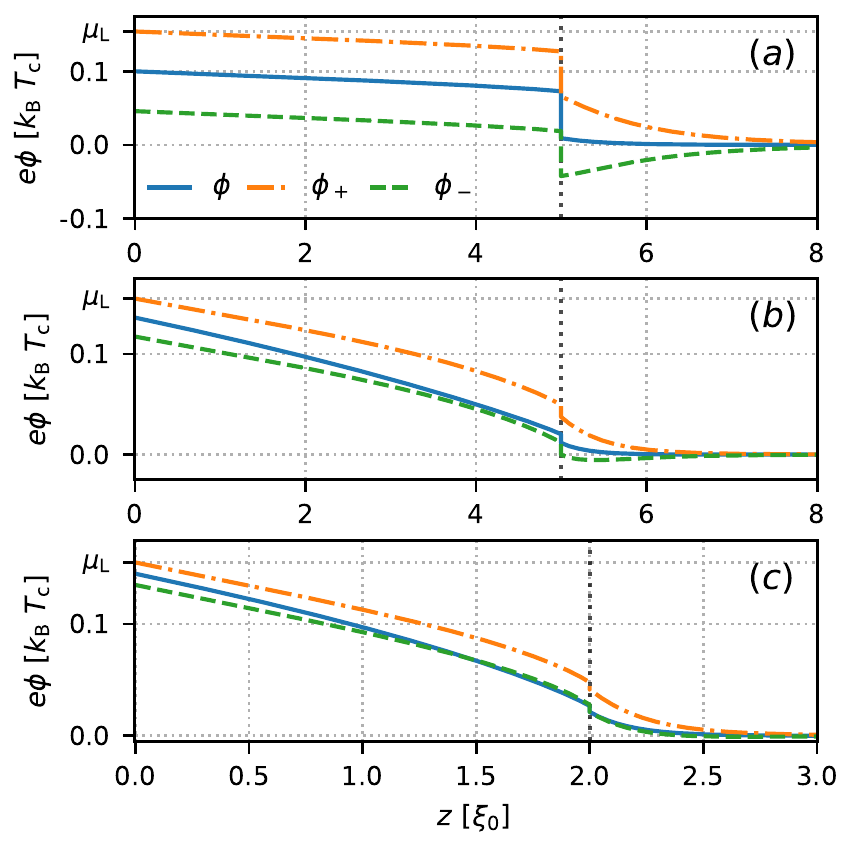}
    \caption{Potential drop in the normal-metal region for a symmetric NISIN system with barrier transparencies $D_\mathrm{L}=D_\mathrm{R}=D=0.8$ at applied bias $\mu_\mathrm{L}=-\mu_\mathrm{R}=eV/2 = 0.155\kb\tc$ for mean free paths and system sizes
    (a) $\mfp= 10\xi_0$, $L=100\xi_0$, $L_\mathrm{N}=5\xi_0$
    (b) $\mfp=\xi_0$, $L=100\xi_0$, $L_\mathrm{N}=5\xi_0$  and
    (c) $\mfp=0.1\xi_0$, $L=25\xi_0$, $L_\mathrm{N}=2\xi_0$.
    Here we display the left normal metal piece and parts of the superconductor. The dotted black vertical lines mark the position of the NIS interface.}
    \label{fig:PotentialComparison_normalMetal}
\end{figure}

\begin{figure}[t]
    \centering
    \includegraphics[scale=1.0]{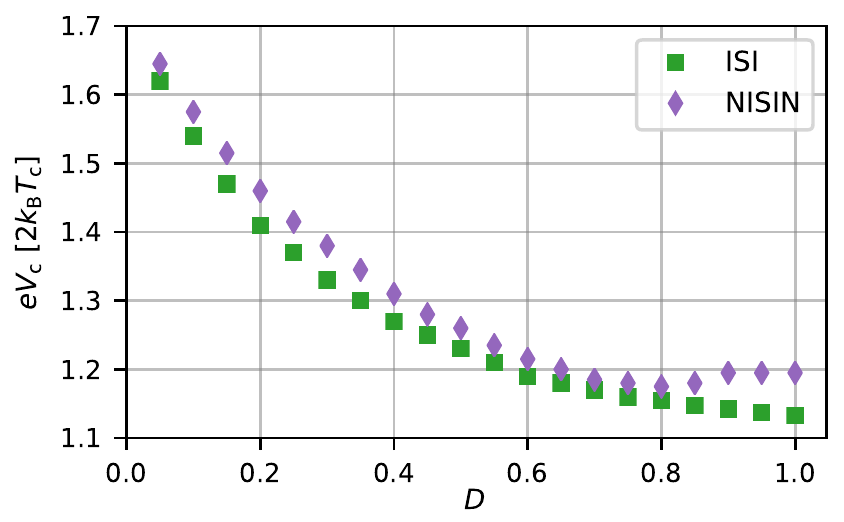}
    \caption{Comparison of the critical voltage for the ISI and NISIN system for intermediate mean free path, $\mfp=\xi_0$. The normal regions in the NISIN system have lengths $L_\mathrm{N}=5\xi_0$, while the superconductor has $L=100\xi_0$. An additional spin-flip mean free path of $\mfp_\mathrm{sf}=10\xi_0$ has been included in the normal metal to let the proximity effect decay away from the interface to the superconductor.}
    \label{fig:CriticalVoltage_NISINvsISI}
\end{figure}

\begin{figure}[t]
    \centering
    \includegraphics[scale=1.0]{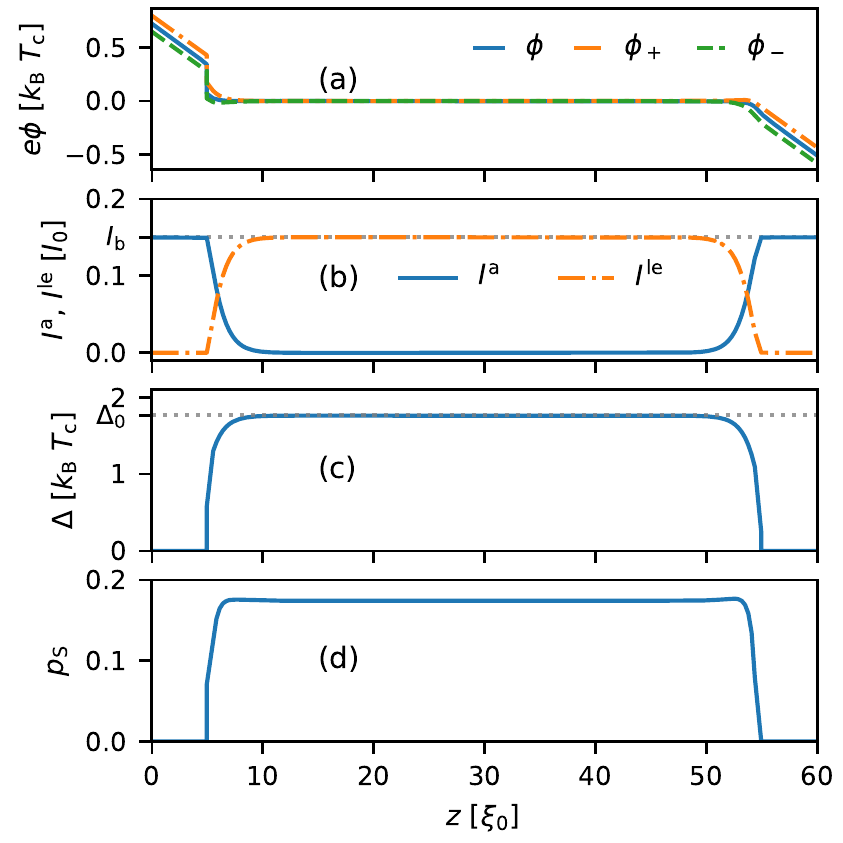}
    \caption{Main observables in the asymmetric NISN system with left and right barrier transparencies $D_\mathrm{L}=0.5$ and $D_\mathrm{R} = 1$. The mean free path is $l_s = \xi_0$. The enforced boundary current is $I_\mathrm{b} = 0.15 I_0$.
    (a) Left-mover and right-mover quasi-potentials $\phi_\pm$ and chemical potential $\phi$, (b) anomalous and local-equilibrium currents, (c) order parameter $\Delta_0$, and (d) superfluid momentum $p_\mathrm{s}$.}
    \label{fig:CurrentBiasedNSN_observables}
\end{figure}

In this section we include the effect of the proximity effect in the normal metal sides. This means that we examine what happens when the central superconducting region is not immediately connected to perfect reservoirs or perfectly ballistic leads, but those reservoirs are located at some distance away from the barriers. We study first a symmetric system, with equal transmissions of the two insulating barriers connecting the superconductor to the normal-metal regions, $D_\mathrm{L}=D_\mathrm{R}=D$, and consider a symmetric bias $\mu_\mathrm{L} = - \mu_\mathrm{R} = eV/2$.

In Fig.~\ref{fig:PotentialComparison_normalMetal} the potential drops in the normal-metal regions are shown for different mean free paths. In the ballistic case, Fig.~\ref{fig:PotentialComparison_normalMetal}(a), the lead is only weakly proximitized and the potential drop is almost linear. For intermediate to small mean free paths, the linear potential drops are increasingly altered by the presence of the proximity effect. The interior of the superconductor is at this applied voltage at ground potential. The enhanced proximity effect in the normal metals reduces the local potential $\phi(z)$ near the N-I-S interface as compared to the normal case, i.e., the N-I-N system studied in Section~\ref{sec:normal}. As a consequence, the proximity effect leads to a larger voltage drop in the normal metal. 

Fig.~\ref{fig:CriticalVoltage_NISINvsISI} shows a comparison of the critical voltages for the NISIN system and the corresponding ISI system. 
For transparencies $D \lesssim 0.75$, we see a slight, almost constant increase of the critical voltage in the NISIN system. The normal ''lead`` gives rise to an additional resistance compared to the ISI case where only the interface resistance determines the current flow. 

At higher transparencies, $D \gtrsim 0.8$, the critical voltage starts to increase again in the NISIN system and plateaus for $D \gtrsim 0.9$, while it monotonically decreases in the ISI case. In this limit, the resistance of the normal lead dominates over the small interface resistance. For this set of parameters, the behavior changes at $D \approx 0.75$. In general, the turning point will depend on the lead resistance compared to the interface resistance and will thus depend on the length of the normal lead and the mean free path.

\begin{figure}[t]
    \centering
    \includegraphics{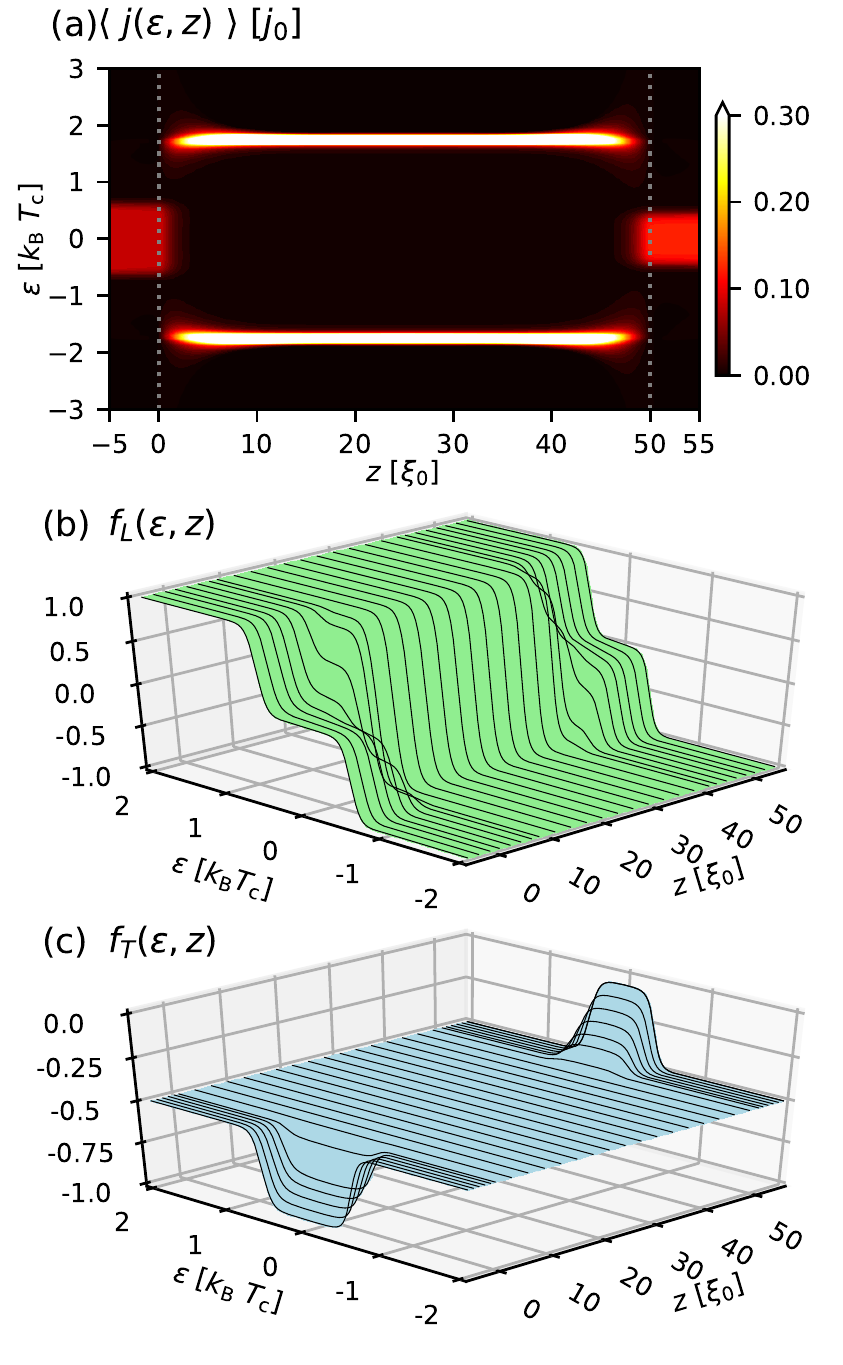}
    \caption{(a) Spectral current density $\langle j(\varepsilon, z)\rangle $ for a normal metal-insulator-superconductor-normal metal (NISN) system with asymmetric interface transparencies ($D_L=0.5, D_R=1)$ and intermediate mean free path $\ell=\xi_0$. The conversion from quasiparticle current in the normal metals carried by states near the Fermi energy to superflow carried by the coninuum states is clearly seen. (b) Energy mode $f_L$ of the distribution function for the same system. The mode is zero for energies less than $eV_\mathrm{L/R}$ in the N region and relaxes to the equilibrium shape deep in the superconductor. (c) The charge mode $f_T$ of the distribution function for the same system.}
    \label{fig:abstractFigure}
\end{figure}

Lastly, we assume that the NISIN structure is not symmetric but rather has different interface transparencies on the two sides. In this case, the potential profile is not symmetric and we must use the current-bias scheme introduce in Sect.~\ref{sec:currentBC}. We specify a boundary current $\Ib$ and find the potentials $\phi_\pm^\mathrm{b}$ that have to be enforced at the system edges. Fig.~\ref{fig:CurrentBiasedNSN_observables} shows an example result of one such calculation. As can be seen, the potential applied on the two sides will be different depending on the interface transparencies. This leads to spectral currents of different widths $2\phi_\pm^\mathrm{b}$ in the normal-region leads, as shown in Fig.~\ref{fig:abstractFigure}(a). The asymmetry between left and right interfaces are also reflected in the distribution functions, see Fig.~\ref{fig:abstractFigure}(b)-(c). Still, in the bulk interior, the distributions have decayed to equilibrium forms and the current is only due to spectral rearrangements due to $p_\mathrm{s}$.

\section{Summary}

In this paper, we studied charge transport through mesoscopic, normal metal-superconductor wires using quasiclassical theory. Such hybrid systems with arbitrary mean free paths were considered, extending the fully ballistic or fully diffusive limits studied in literature. We performed charge-current conserving calculations and studied phenomena such as charge imbalance, the conversion of quasiparticle current to superflow, and the critical voltage of the superconductor.

For normal metal-superconductor interfaces that are not of pinhole-type, we observed a critical voltage $V_\mathrm{c}$ at which the superconductor turns normal. This connects with a similar observation for the fully-diffusive system with a fully-transparent interface \cite{Keizer2006Apr}. The critical voltage was found to be the result of an interplay between Doppler shifts and the injected nonequilibrium energy mode. We investigated the influence of  both the mean free path and interface transparency on the critical voltage. 
In general, $V_\mathrm{c}$ increased with smaller interface transparency as this reduced the current through the structure. The effect of impurities on the critical voltage was found to depend on the interface transparency. Higher impurity concentration increased $V_\mathrm{c}$ for very transmissive interfaces but reduced it for all transparencies below $D = 0.75$. We obtained similar results for superconductors connected on both ends to normal-metal leads. However, we find additional oscillating solutions in the ballistic regime of $\ell > \xi_0$, which led to an increased critical voltage in such systems. 
Lastly, short pieces of normal-metal leads were included to study the influence of the proximity effect on transport. We found only small corrections for weakly-proximitized ballistic systems. For intermediate to diffusive systems, the additional resistance of the normal lead increased the critical voltage compared to when the proximity effect is neglected.

It should be noted, that the estimate of $V_c$ from our calculations are obtained neglecting many phenomena that may become of importance near breakdown, such as for instance the electron-phonon interaction. It would be of interest to improve the model and consider the highly non-equilibrium distribution function in Fig.~\ref{fig:DeltaKernelFigureL1} and its influence on inelastic processes, but this is beyond the scope of the present paper.

The method introduced in this paper is applicable to any kind of mesoscopic superconducting systems for arbitrary mean free path. It can thus be used to study other phenomena, such as spin-charge density separation \cite{shevtsov_spin_2014,beckmann_spin_2016}, heat flow \cite{Richard2016}, or unconventional superconductors with other order-parameter symmetries \cite{kashiwaya_tunnelling_2000,lofwander_andreev_2001}, where the surface and interface physics may be non-trivial even in the equilibrium state \cite{Hakansson2015,Holmvall2018,Holmvall2020}.

\begin{acknowledgments}
We thank the Swedish research council for financial support.
\end{acknowledgments}


\providecommand{\noopsort}[1]{}\providecommand{\singleletter}[1]{#1}%

\end{document}